\documentclass[sigconf, nonacm]{acmart}

\newcommand\vldbdoi{XX.XX/XXX.XX}
\newcommand\vldbpages{XXX-XXX}
\newcommand\vldbvolume{14}
\newcommand\vldbissue{1}
\newcommand\vldbyear{2020}
\newcommand\vldbauthors{\authors}
\newcommand\vldbtitle{\shorttitle} 
\newcommand\vldbavailabilityurl{https://github.com/puddingfjz/vllm}
\newcommand\vldbpagestyle{plain}

\newtheorem{definition}{Definition}
\newtheorem{theorem}{Theorem}
\usepackage[linesnumbered]{algorithm2e}
\RestyleAlgo{ruled}
\SetKwComment{Comment}{/* }{ */}

\usepackage{cleveref}
\usepackage{amsfonts}

\usepackage{subcaption}
\usepackage{booktabs} %

\newcommand{\ours}{Ours~}

\begin{document}
\title{Improving the End-to-End Efficiency of Offline Inference for Multi-LLM Applications Based on Sampling and Simulation}

\author{Jingzhi Fang}
\affiliation{%
  \institution{HKUST}
}
\email{jfangak@connect.ust.hk}

\author{Yanyan Shen}
\affiliation{%
  \institution{Shanghai Jiao Tong University}
}
\email{shenyy@sjtu.edu.cn}

\author{Yue Wang}
\affiliation{%
  \institution{Shenzhen Institute of Computing Sciences}
}
\email{ywangby@connect.ust.hk}

\author{Lei Chen}
\affiliation{%
  \institution{HKUST, HKUST(GZ)}
}
\email{leichen@cse.ust.hk}

\begin{abstract}
As large language models (LLMs) have shown great success in many tasks, they are used in various applications. While a lot of works have focused on the efficiency of single-LLM application (e.g., offloading, request scheduling, parallelism strategy selection), multi-LLM applications receive less attention, particularly in offline inference scenarios.
In this work, we aim to improve the offline end-to-end inference efficiency of multi-LLM applications in the single-node multi-GPU environment.
The problem involves two key decisions: (1) determining which LLMs to run concurrently each time (we may not run all the models at the same time), and (2) selecting a 
parallelism strategy to use for each LLM. 
This problem is NP-hard. 
Naive solutions may not work well because the running time for a model to complete a set of requests depends on the request workload and the selected parallelism strategy, and they lack an accurate model of the running time.
As the LLM output lengths are unknown before running, to estimate the model running time, we propose a sampling-then-simulation method which first estimates the output lengths by sampling from an empirical cumulative function we obtained from a large dataset in advance, and then simulates the LLM inference process accordingly.
Based on the simulation, we estimate the per-iteration latencys to get the total latency.
A greedy method is proposed to optimize the scheduling of the LLMs in the application across the GPUs.
We then propose a framework SamuLLM which contains two phases: planning, which calls the greedy method for an application and running, which runs the application and dynamically adjust the model scheduling based on the runtime information.
Experiments on 3 applications and a mixed application show that SamuLLM can achieve 1.0-2.4$\times$ end-to-end speedups compared to the competitors.

\end{abstract}

\maketitle

\pagestyle{\vldbpagestyle}
\begingroup\small\noindent\raggedright\textbf{PVLDB Reference Format:}\\
\vldbauthors. \vldbtitle. PVLDB, \vldbvolume(\vldbissue): \vldbpages, \vldbyear.\\
\href{https://doi.org/\vldbdoi}{doi:\vldbdoi}
\endgroup
\begingroup
\renewcommand\thefootnote{}\footnote{\noindent
This work is licensed under the Creative Commons BY-NC-ND 4.0 International License. Visit \url{https://creativecommons.org/licenses/by-nc-nd/4.0/} to view a copy of this license. For any use beyond those covered by this license, obtain permission by emailing \href{mailto:info@vldb.org}{info@vldb.org}. Copyright is held by the owner/author(s). Publication rights licensed to the VLDB Endowment. \\
\raggedright Proceedings of the VLDB Endowment, Vol. \vldbvolume, No. \vldbissue\ %
ISSN 2150-8097. \\
\href{https://doi.org/\vldbdoi}{doi:\vldbdoi} \\
}\addtocounter{footnote}{-1}\endgroup

\ifdefempty{\vldbavailabilityurl}{}{
\vspace{.3cm}
\begingroup\small\noindent\raggedright\textbf{PVLDB Artifact Availability:}\\
The source code, data, and/or other artifacts have been made available at \url{\vldbavailabilityurl}.
\endgroup
}

\section{Introduction}~\label{sec: intro}
Large language models (LLMs) have shown their powerful capability in many tasks, therefore, they have been used in various applications.
The applications include not only single-LLM applications, such as chatbots~\cite{dam2024complete}, personal assistants~\cite{li2024personal}, and various domain-specific LLM-based tools~\cite{joel2024survey}, but also emerging multi-LLM applications, e.g., LLM ensembling~\cite{Blender}, LLM routing~\cite{hu2024routerbench}, LLM-based document summarization followed by evaluation~\cite{BooookScore}, etc.
In terms of the request arrival mode, the LLM-based applications can be divided into 2 groups: offline scenario applications, whose requests are provided in advance (e.g., benchmarking~\cite{liang2022holistic}, information extraction~\cite{narayan2018don}, and data
wrangling~\cite{narayan2022can}), and online scenario applications (e.g., chatbots~\cite{dam2024complete}) whose requests may arrive at any time.
The inference efficiency of LLM applications is critical for saving the application running expense (e.g., renting an H100 on Amazon takes about 4 USD per hour~\cite{Amazonprice}) and accelerating the AI development cycle.
While people have done a lot of work to improve the inference efficiency of single-LLM applications~\cite{kwon2023efficient,yu2022orca,sheng2023flexgen,zheng2024sglang}, multi-LLM applications receive less attention, particularly in offline inference scenarios.
In this paper, we focus on multi-LLM applications in the offline scenario and try to improve their overall inference throughput.

\begin{figure}[t]
     \centering
     \includegraphics[width=1\linewidth]{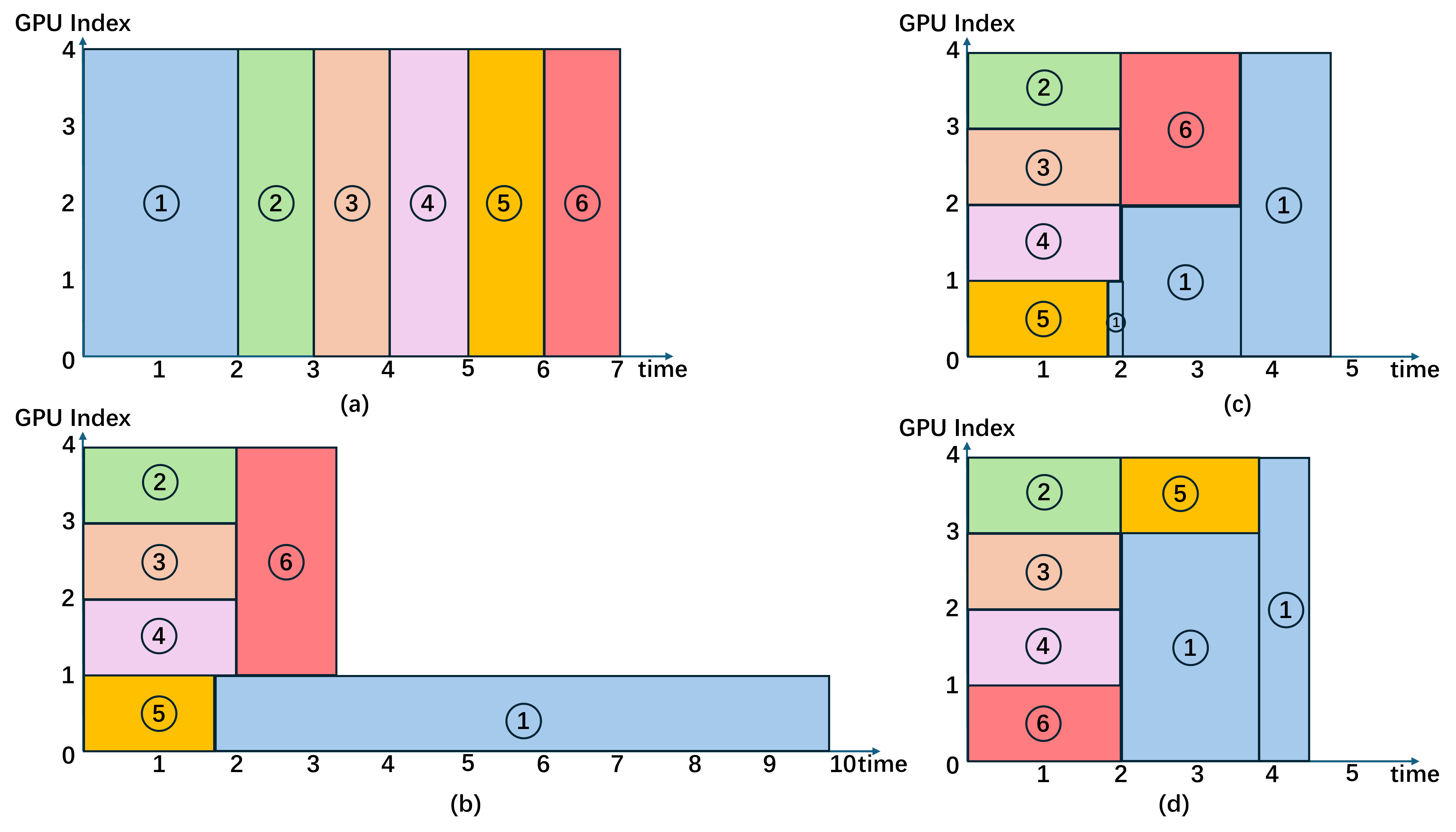}
     \caption{Different application running plans result in different end-to-end efficiency. The running time of a model does not necessarily decrease linearly with the number of GPUs assigned to it. This figure shows 4 different execution plans for 6 LLMs in an application with no dependency among the models: (a) running LLMs sequentially and assigning all GPUs to the running LLM each time; (b) running as many LLMs concurrently as possible and splitting the GPUs among the LLMs event (model preemption is not allowed); (c) the same as (b) but allowing model preemption; (d) assigning more GPUs to model 1 in the second stage to achieve higher throughput, rather than always partitioning the GPUs evenly.}
     \label{fig: exec plan matters}
\end{figure}

Specifically, this work focuses on the single-node multi-GPU environment, where multiple models can run in parallel and each model's inference can be distributed as well. 
In this kind of environment, the common practice of distributed inference for an LLM involves data parallelism (partitioning input requests across replicated model instances) and tensor parallelism (distributing the computation of individual model layers across multiple GPUs)~\cite{vllmParallelPractice}.
Therefore, we call the combination of a specific data parallelism setting and a specific tensor parallelism setting for a model a \textbf{model execution plan} in this paper.

As there may be more models in an application than the available GPUs, determining how to schedule the multi-LLM application on the available GPUs to maximize the overall throughput requires the joint optimization of (1) the selection of the subset of models to run concurrently each time and (2) the corresponding model execution plans.
This problem is challenging and the hardness comes from several aspects.
(1) The time for a model to complete a set of requests depends on the request workload (the request input lengths and output lengths) and the model execution plan. 
The best throughput of different model execution plans does not necessarily increase linearly with the GPU number, since adding more GPUs can result in insufficient workload per GPU, leading to computation resource underutilization. 
For example, it takes chatglm3-6b~\cite{glm2024chatglm} 48s to complete 1000 requests on 1 A100 GPU and 32s on 8 A100 GPUs.
(2) For different models, the relationships between the best throughput of different model execution plans and the GPU number are different.
(3) The number of running requests for a model may change significantly during inference, resulting in varying throughput even with the same distributed inference strategy, e.g., when summarizing documents of different lengths chunk by chunk, only chunks from the long documents are left as inference progresses (\Cref{fig:computation graphs} (d)).
(4) The request output lengths are unknown before the application runs, i.e., we do not know the exact request workloads.

Naive methods may not work well on this problem.
For instance, consider 4 GPUs and an LLM ensembling application with 6 models, which generate answers for the same requests independently. 
Suppose for model 1, the best throughput increases linearly with the number of GPUs and the running time using 1 GPU is 8, while for other models, $n\times$ more GPUs bring less than $n\times$ throughput improvement.
One naive solution is to run the models one by one, each using 4 GPUs (\Cref{fig: exec plan matters} (a)).
We can also run as many models concurrently as possible each time and split the GPUs evenly among them, as shown by~\Cref{fig: exec plan matters} (b), where a model cannot change its distributed inference strategy once started.
However, in~\Cref{fig: exec plan matters} (b), 3 GPUs become idle after model 6 finishes, causing significant GPU resource waste.
If model preemption is allowed and a model's execution plan can be changed, the total running time can be reduced a lot by making the GPUs always occupied (\Cref{fig: exec plan matters} (c)).
We can further improve the overall throughput by assigning more GPUs to model 1 in the second stage (where model 5 and model 1 runs concurrently), rather than partitioning the GPUs evenly (\Cref{fig: exec plan matters} (d)), because the throughput increases linearly as the number of GPUs for model 1, not for model 5.

Therefore, to schedule a multi-LLM application with the maximum overall throughput (i.e., minimum overall latency), we are faced with 2 challenges:
(1) estimating the running time of a model given a set of requests and a model execution plan accurately,
and (2) efficiently searching for the minimum-total-latency scheduling of the application on the GPUs, where model preemption is allowed.

To overcome the first challenge, the biggest obstacle is that the request output lengths of an LLM are unknown before it runs, as an LLM generates an output auto-regressively.
Existing works try to predict the LLM output lengths directly~\cite{jin2023s,hu2024inference,stojkovic2024dynamollm}, e.g., training an LLM to predict LLM output lengths, or ask LLMs to generate the output lengths via prompting~\cite{zheng2023response}.
However, this kind of approach has two disadvantages: first, the prediction may not be accurate; second, querying a model, especially an LLM, for the request output lengths will introduce additional running time cost.
On the other hand, since we aim to estimate the time for a model to complete a set of requests rather than the time to complete each individual request, we do not need to know the exact output length of each request for an LLM. 
Instead, it is sufficient for us to know how the output lengths of these requests are distributed as a whole.
In fact, we have observed that the output length of an LLM generally follows a distribution irrelevant to the concrete request,
except that there are some special instructions in the request or in the inference settings (e.g., asking the model to only provide the answer to a multiple-choice question or setting the maximum output length limit to 900).
Therefore, we can obtain the empirical cumulative distribution functions (eCDFs) in advance by collecting the output lengths of a large set of requests.
Then, we can randomly sample the output lengths for the given requests under the eCDFs. 
The latency of one iteration computation (one forward pass) of an LLM can be estimated easily given the running request information (e.g., the number of requests and their lengths) in that iteration, using a set of linear functions.
Based on the sampled output lengths, we can simulate the inference process to estimate the running request information in each iteration for an LLM.
Then, based on the per-iteration latency estimation, we can finally get the estimated total inference time.

\begin{figure*}[t]%
    \centering
    \begin{subfigure}{0.32\textwidth}
        \centering
        \includegraphics[width=\linewidth]{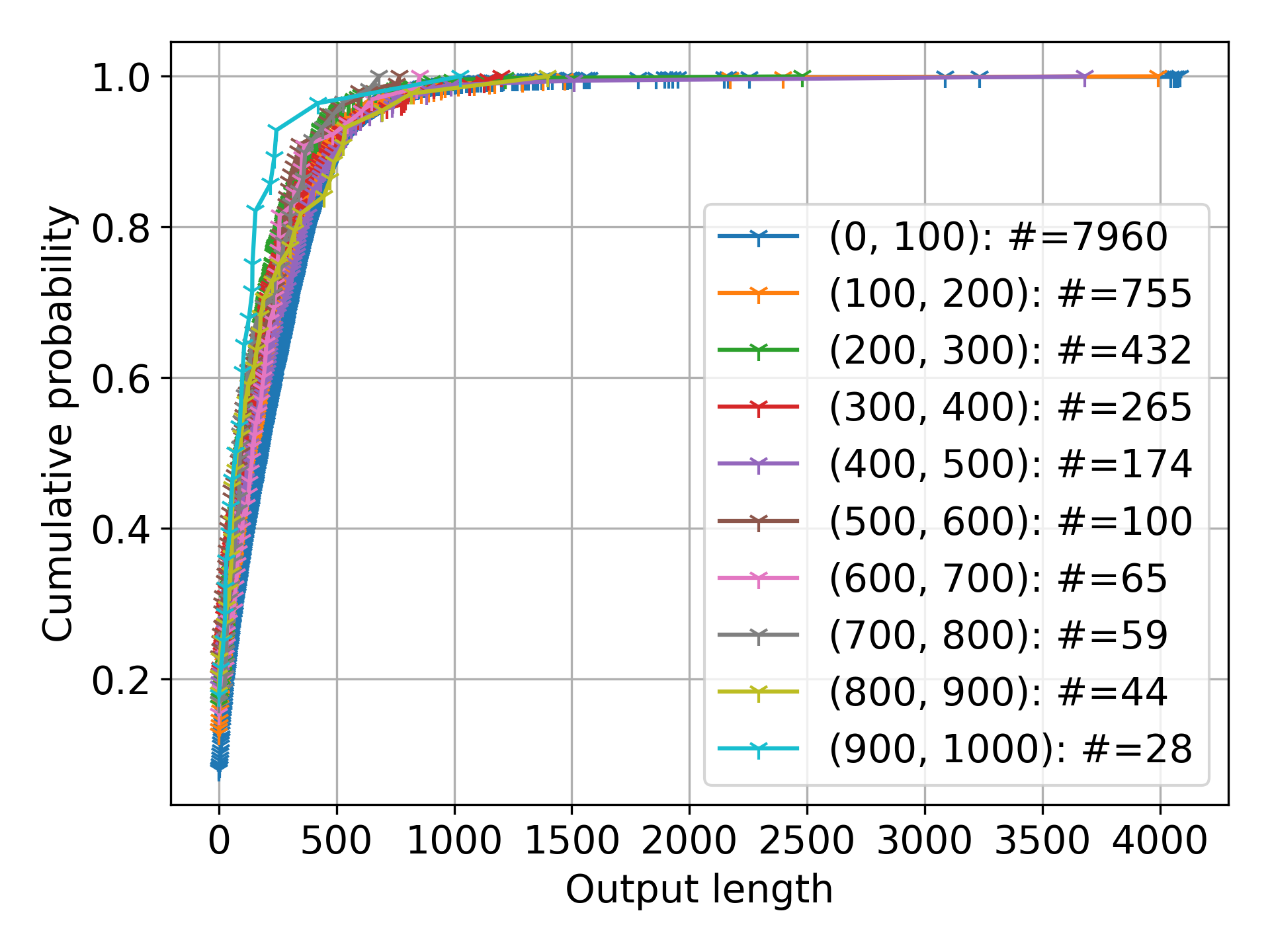} %
        \caption{}
        \label{fig:ecdf1}
    \end{subfigure}
    \begin{subfigure}{0.32\textwidth}
        \centering
        \includegraphics[width=\linewidth]{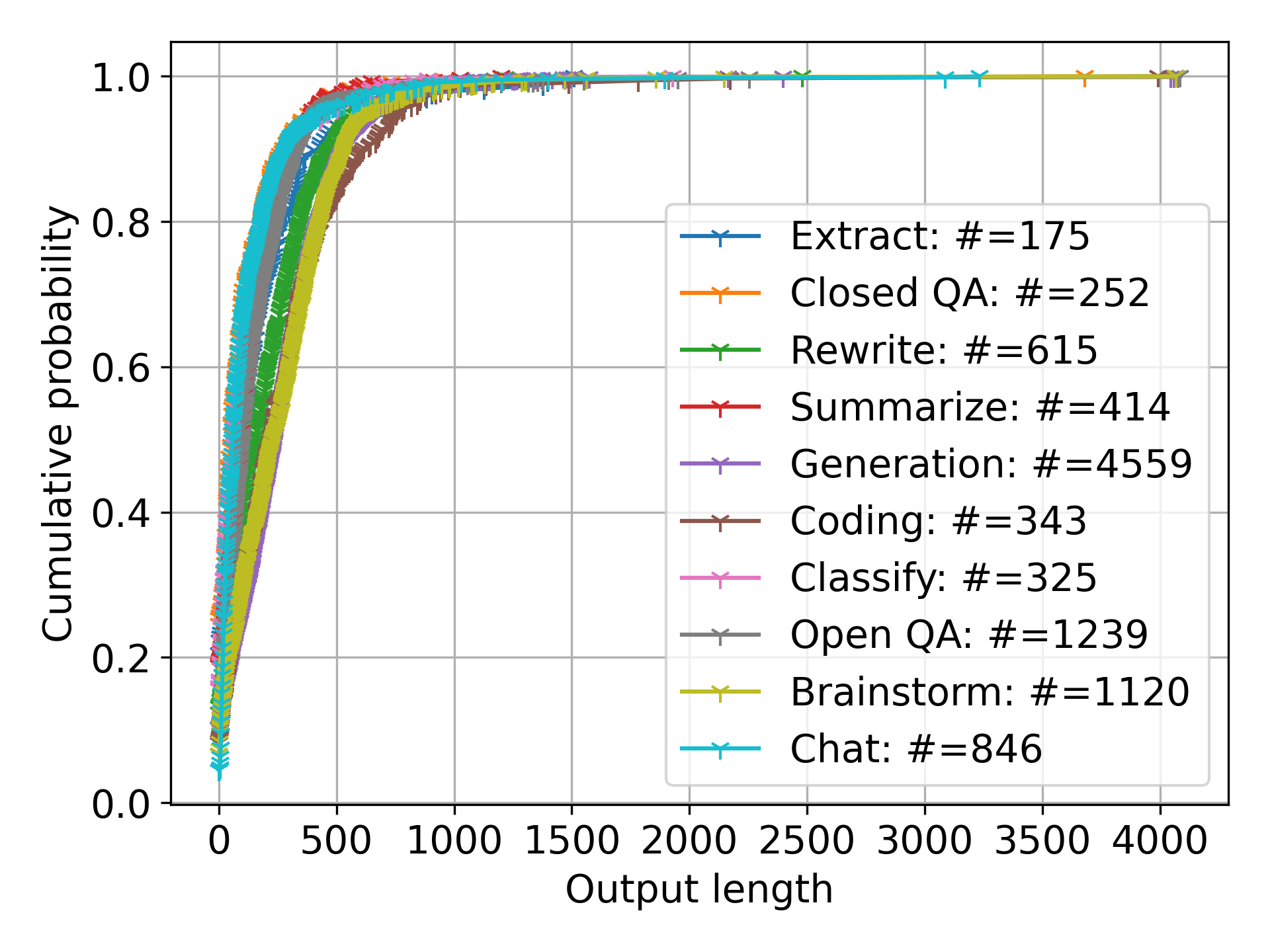} %
        \caption{}
        \label{fig:ecdf2}
    \end{subfigure}
    \begin{subfigure}{0.32\textwidth}
        \centering
        \includegraphics[width=\linewidth]{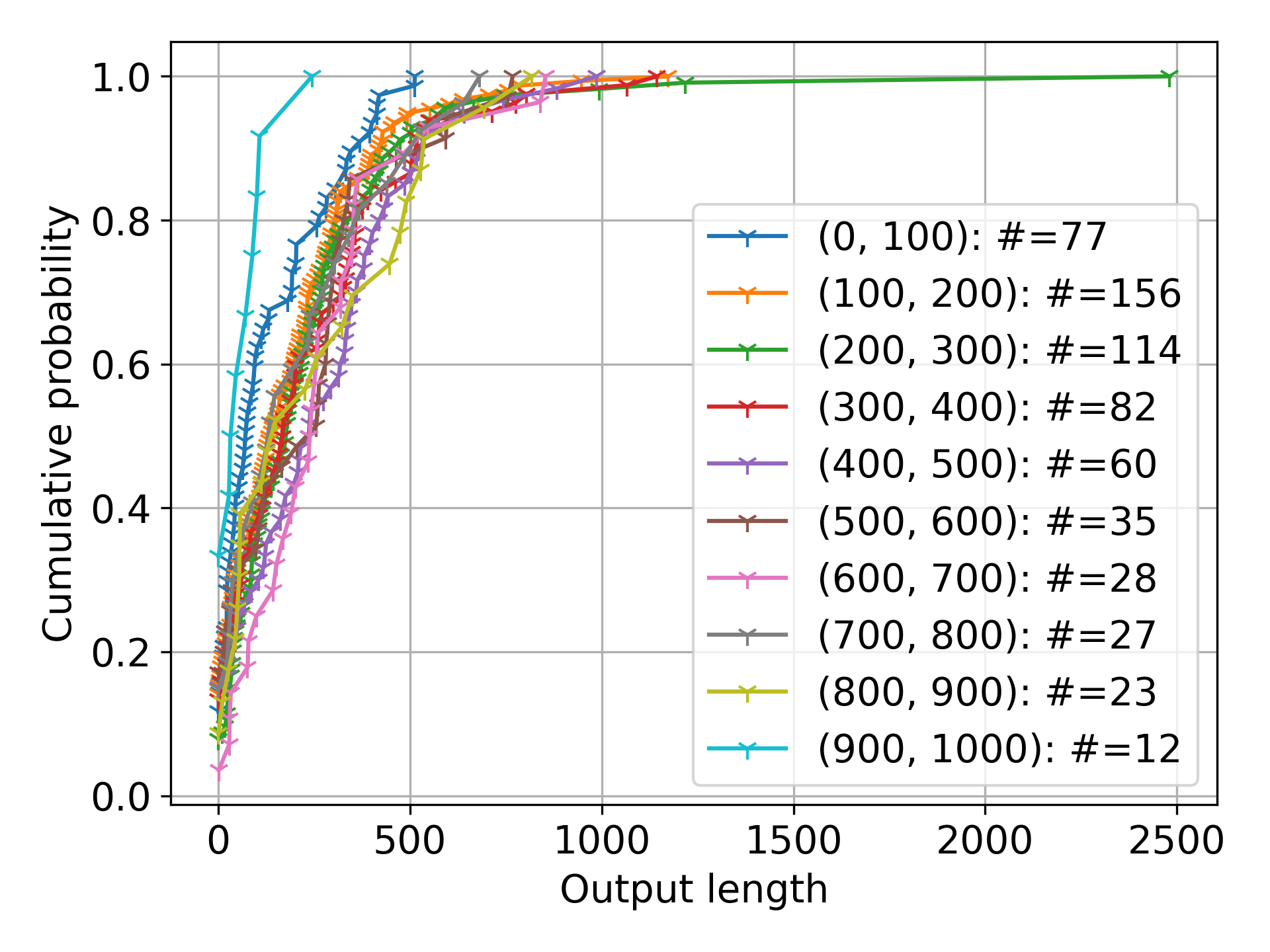} %
        \caption{}
        \label{fig:ecdf3}
    \end{subfigure}
    \caption{Output length eCDFs: (a) eCDFs of different input length regions; (b) eCDFs of different request categories; (c) eCDFs of different input length regions in the ``Rewrite'' category.}
    \label{fig:outlen distribution}
\end{figure*}

With the sampling-then-simulation cost model, we try to search for a high-throughput scheduling for the application on the GPUs.
We call the time period where a set of models run concurrently with their execution plans unchanged an \textbf{execution stage}.
The application running process may consist of multiple execution stages, as shown by~\Cref{fig: exec plan matters}, where time 0-2 in~\Cref{fig: exec plan matters}(d) is a stage with model 2,3,4,6 running. 
We can model the inference of a multi-LLM application as a computation graph, where each node represents an LLM and each edge represents the data dependency between LLMs.
A model will be considered to run in an execution stage only if its input models are either finished in any previous stages or running in the same stage (the latter case introduces \textit{model-level pipeline parallelism}).

Determining the minimum-total-latency sequence of execution stages (i.e., the scheduling) for an application given a set of requests and a set of GPUs is NP-hard, because the bin packing problem is a special case of it.
We use a greedy search method for this problem.
The main idea is that, for each stage, we iteratively select the model and the execution plan with the highest per-GPU throughput increase (possible preemption costs are considered) if we assign more GPUs to the model, which is an idea adapted from Optimus~\cite{peng2018optimus}.
We choose the first-model-finish time as the natural boundary of an execution stage to avoid leaving GPUs idle, i.e., once a model in a stage is finished, the stage ends and the next stage starts, where the remaining requests for each model will be updated and both the running models and their execution plans will be reconsidered.
In this way, we determine the execution stages iteratively until all the models in the application are finished.

Based on the cost model and the greedy search method, we implement a framework SemuLLM to run the multi-LLM applications, which contains two phases: the planning phase, which determines the application scheduling, and the running phase, which runs the application according to the scheduling we obtained.
When the application is running, if the order in which the LLMs complete their requests is not as the scheduling estimates (due to the cost model errors), SemuLLM will dynamically adjust the execution stages according to the runtime information without redoing the search.

We conduct experiments on 3 applications and a mixed application to compare SemuLLM with two commonly used heuristics. 
The results show that (1) we can achieve 1.0-2.4$\times$ end-to-end speedup (including the application scheduling search time) compared to the competitors, (2) not allowing preemption would make the throughput 1.0-1.4$\times$ lower, and (3) our cost model can effectively guide us to a good sequence of execution stages for an application.

Our contributions can be summarized as follows:
\begin{enumerate}
    \item We propose a sampling-then-simulation cost model to estimate the running time of an LLM given a set of requests and its execution plan.
    \item We model the multi-LLM application as a computation graph, define the problem of searching for the minimum-total-latency application scheduling (NP-hard), and propose a greedy search method for it.
    \item We implement a framework SemuLLM which runs multi-LLM applications according to the scheduling it finds and supports dynamic execution stage adjustment in case the estimation differs from the actual running process.
    \item We conduct experiments on various multi-LLM applications and validate the effectiveness and the efficiency of SemuLLM.
\end{enumerate}

\section{Cost Model Insight}~\label{sec: motivation and insight}
In this section, we present some insights we obtained from experiments to design the cost model which estimates the running time for a model to complete a set of requests using a specific execution plan.

\textbf{Request output lengths follow a probability distribution.}
Furthermore, this distribution is not very related to the request length or category, except when the request includes instructions about the output length or when the output length limit is set in the inference settings (e.g., asking the LLM to provide only the answer to a multiple-choice question in the request, or forcing the inference to stop after generating 900 tokens in the output).
This insight is obtained from our experiment observation. Specifically, we use the No Robots dataset~\cite{no_robots}, which contains 10 different categories instructions, including Generation, Open QA, Brainstorm, Chat, Rewrite, Summarize, Coding, Classify, Closed QA, Extract.
We randomly sample 10000 requests from the No Robots dataset, send them to different LLMs, and collect their output lengths.
We draw the empirical cumulative distribution functions (eCDFs) of these output lengths.
\Cref{fig:outlen distribution} shows some eCDFs for vicuna-13b-v1.5~\cite{chiang2023vicuna}.
From~\Cref{fig:outlen distribution} we can see that, the eCDFs are similar regardless of the request length (\Cref{fig:outlen distribution}(a)) or the request content category (\Cref{fig:outlen distribution}(b)).
\Cref{fig:outlen distribution}(c) focuses on ``Rewrite'' requests and demonstrates that the eCDFs are similar across different regions of request lengths as well.
Therefore, we construct eCDFs of output lengths for different models using the No Robots dataset~\cite{no_robots} in this work to generate output length estimations by sampling.

\begin{figure}[t]%
    \centering
    \begin{subfigure}{0.3\textwidth}
        \centering
        \includegraphics[width=\linewidth]{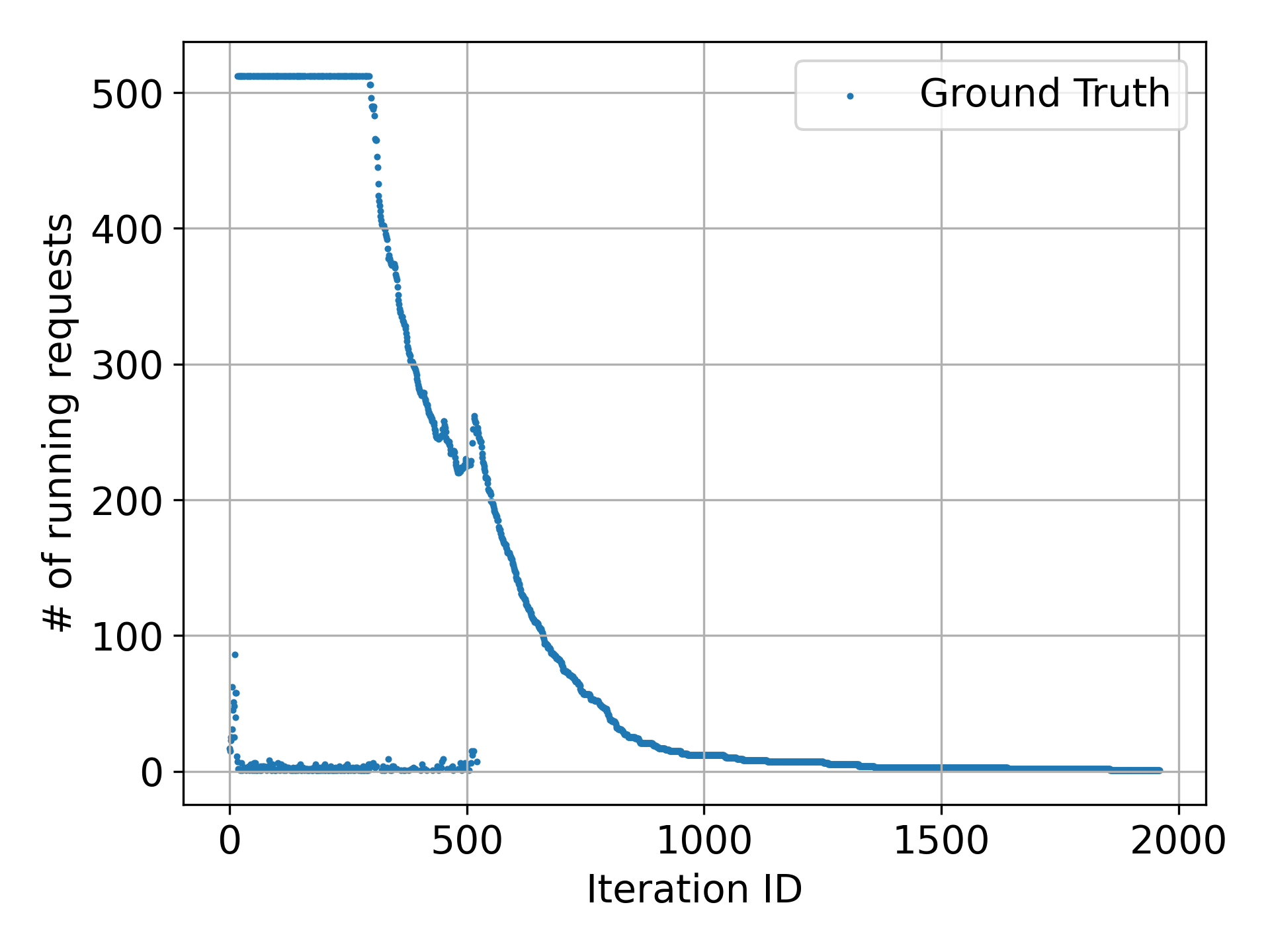} %
        \caption{}
        \label{fig:real}
    \end{subfigure}
    \begin{subfigure}{0.3\textwidth}
        \centering
        \includegraphics[width=\linewidth]{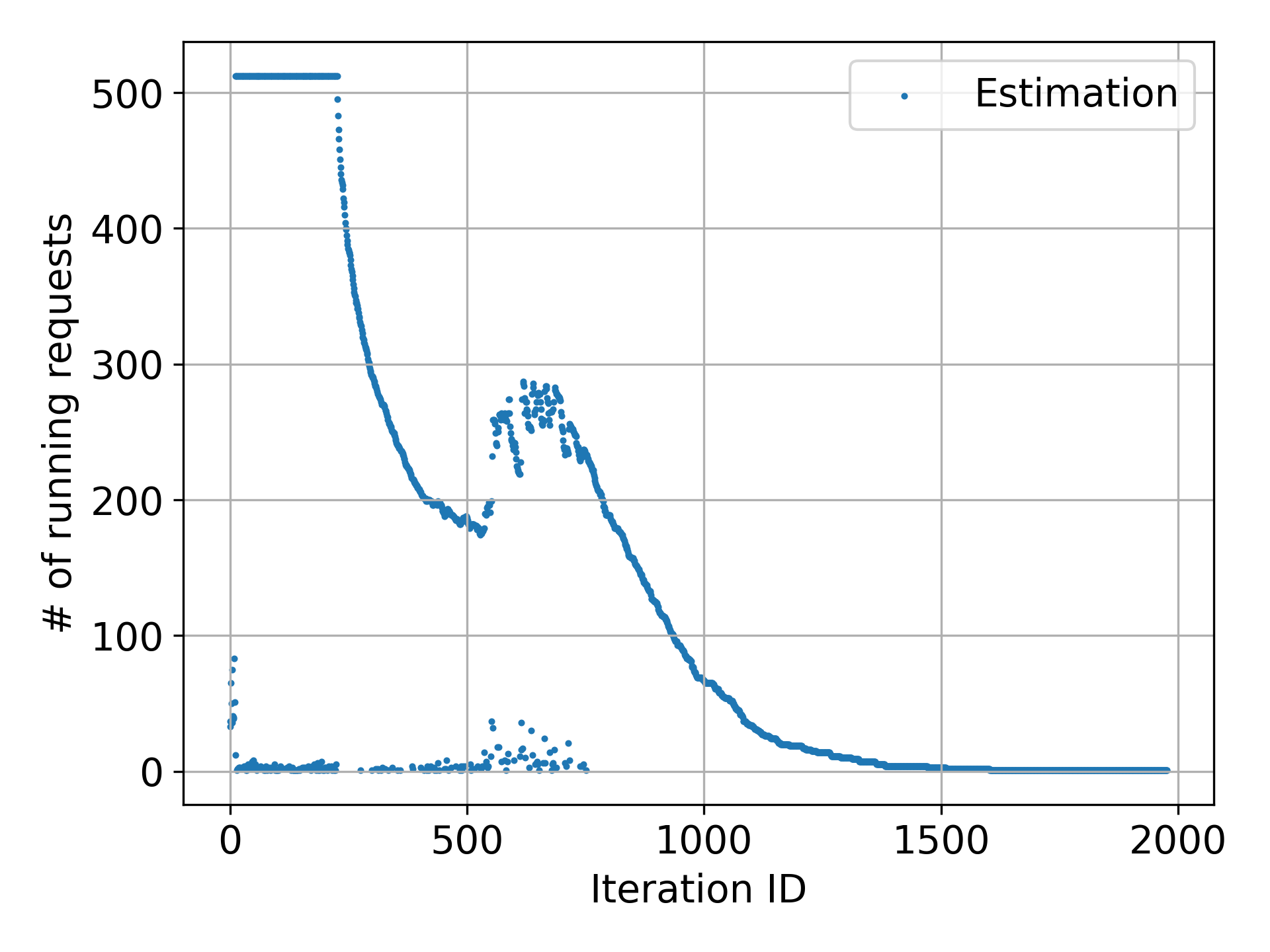} %
        \caption{}
        \label{fig:fake}
    \end{subfigure}
    \caption{Running request number of each iteration in (a) the real inference process and (b) the simulated inference process, respectively.}
    \label{fig:fake scheduling}
\end{figure}

\textbf{The inference process can be simulated.}
Specifically, as long as we know the request scheduling policy of an LLM engine (e.g., the First Come First Serve (FCFS) policy used in vLLM~\cite{kwon2023efficient}), we can simulate the running requests participating in every inference iteration based on the estimated request output lengths.
\Cref{fig:fake scheduling} compares the running request number over inference iterations for vicuna-13b-v1.5~\cite{chiang2023vicuna} to complete 1000 requests from the SharedGPT dataset~\cite{chiang2023vicuna} in the real inference process and the simulated inference process, respectively.
The simulation captures the main curve pattern of the real inference process, despite some differences.
Based on the simulation, if we can estimate the latency of each iteration, we can estimate the overall inference time.

\begin{figure*}[t]
     \centering
     \includegraphics[width=1\linewidth]{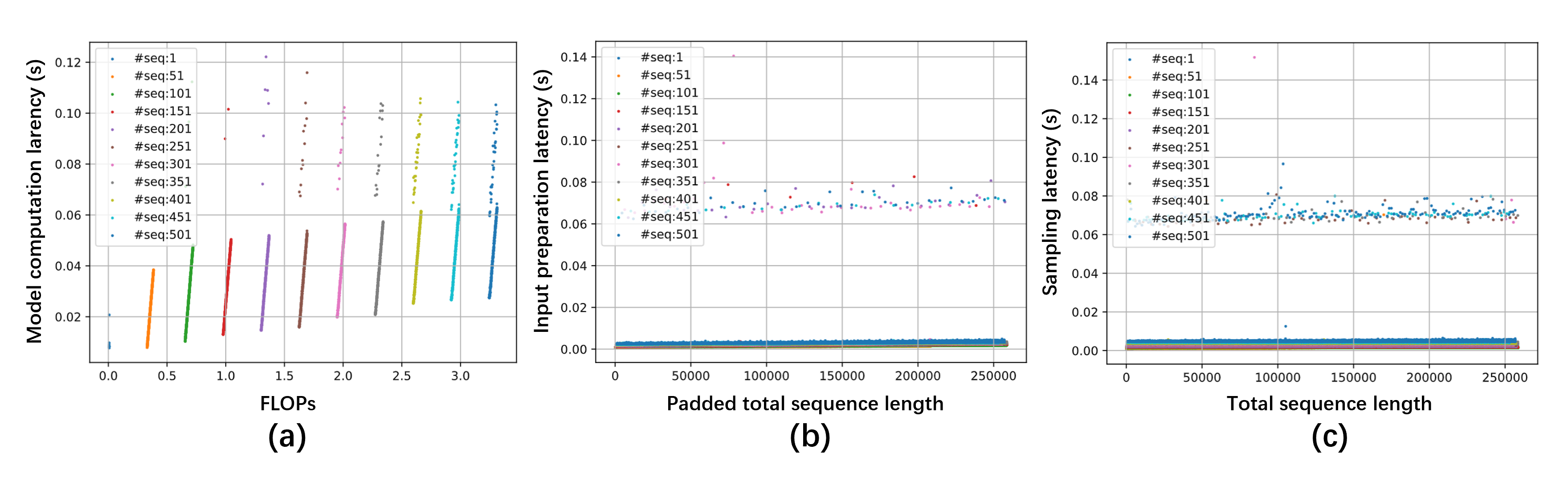}
     \caption{Three major components of the LLM per-iteration latency: (a) model computation latency, (b) input preparation latency, and (c) output token sampling latency.}
     \label{fig:per iter latency}
\end{figure*}

\textbf{The per-iteration inference latency can be estimated using a set of linear functions.}
We profile the latency of each iteration in the inference process of Llama 7B~\cite{touvron2023llama} for a given set of requests, running on 1 GPU with vLLM~\cite{kwon2023efficient}. 
Based on it, we choose to divide the per-iteration latency into three parts to model the main contributing factors:
(1) model computation (computing the results of all model layers), (2) input preparation before model computation, and (3) output token sampling after model computation.

The model computation latency is related to the number of running requests and the number of FLOPs of the iteration, which can be estimated by 
\begin{equation}~\label{eq: prefill flops}
    \mathsf{FLOPs}_{\mathsf{prefill}} = L(c\cdot Bs + 2Bhs^2/\mathsf{tp})
\end{equation}
\begin{equation}~\label{eq: decode flops}
    \mathsf{FLOPs}_{\mathsf{decode}} = L(c\cdot B + 2hS/\mathsf{tp})
\end{equation}
\Cref{eq: prefill flops,eq: decode flops} compute the FLOPs in a prefill iteration and the FLOPs in a decode iteration, respectively, where $L$ is the number of layers, $B$ is the number of requests, $s$ is the maximum request length of all running requests, $h$ is the hidden dimension, $\mathsf{tp}$ is the tensor parallelism degree, $S$ is the total request length, and $c$ is a constant computed by summing up the sizes of all the model weight matrices that are related to matrix multiplication operations.
\Cref{fig:per iter latency} (a) presents our profiling results about the per-iteration model computation latency (there are some noise points sparsely distributed, and we can ignore them). It shows that, given a request number $B$ ($\#\mathsf{seq}$ in the figure), this part of latency increases linearly with the number of FLOPs. 

Besides, we find the per-iteration input preparation latency is related to the request number $B$ and the padded total length of the running requests ($B\cdot s$), while the per-iteration sampling latency is related to $B$ and the total length of the running requests without padding ($S$), respectively.
\Cref{fig:per iter latency} (b,c) show that we can also use linear functions associated with different request numbers $B$ to model these two parts of latency (there are also some noise points in the upper part of the figures).

To estimate the total running time for a model to complete a set of requests, we also need to consider the time of model loading, which includes loading the model weights on specific GPUs and preparing the communication environment if tensor parallelism is used.
We can profile the model loading time for different models and different execution plans to build a cost table in advance.

Based on the simulation in~\Cref{fig:fake scheduling} (b), by estimating the per-iteration latency, we estimate the time for vicuna-13b-v1.5~\cite{chiang2023vicuna} to complete the 1000 requests from SharedGPT~\cite{chiang2023vicuna} is 98 seconds, while the real running time we measured is 92 seconds, with an error ratio of $6.5\%$.
Considering the model loading time, our estimation is 128 seconds, while the real running time is $127$ seconds, which is almost the same as the estimation.

\section{Problem Formulation}~\label{sec: problem formulation}

\begin{figure}[t]
     \centering
     \includegraphics[width=0.8\linewidth]{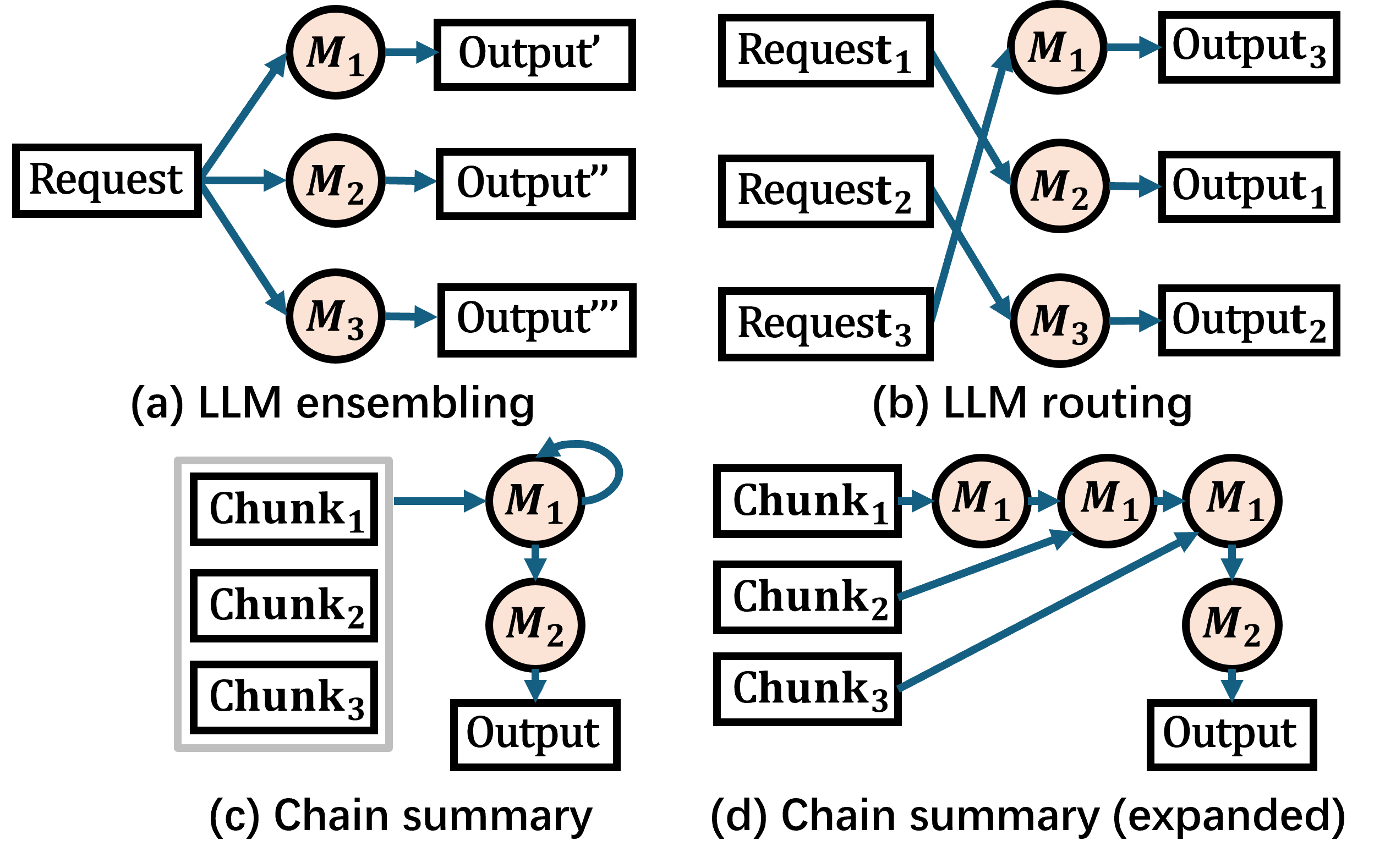}
     \caption{The computation graphs of different applications.}
     \label{fig:computation graphs}
\end{figure}

In this section, we first introduce some concepts used in this work.
We then define the optimization problem and prove its hardness.

\textbf{Computation graph.} 
As shown by~\Cref{fig:computation graphs}, we can model a multi-LLM application as a computation graph, where each node corresponds to an LLM and each edge corresponds to a data flow between LLMs.
An edge may point from one node to another node, or from one node to itself.
For example, in Chain Summary (\Cref{fig:computation graphs} (c)), a long document is split into multiple chunks, and the summary will be updated as we step through the document chunk-by-chunk~\cite{BooookScore} (done by the summarization node $M_1$).
We can also expand the self-loops in the computation graph to make it acyclic, as shown in~\Cref{fig:computation graphs} (d), in which the number of the summarization nodes equals the number of chunks.

An LLM may have multiple input sources in the application.
Some LLMs need to concatenate the input from different sources to get a complete input request; some take inputs from different sources as individual requests; an LLM may also not take all the output from a model as input.
For example, in~\Cref{fig:computation graphs} (d), each $M_1$ node updates the summary based on both the input chunk and the old summary; for the evaluator node $M_2$ which evaluates the final summary quality, the inputs are from different $M_1$ nodes if there are documents of different lengths; for each $M_1$ node, only the output that is the final summary of a document will be the input of $M_2$.

\textbf{Model execution plan.}
We consider two types of parallelism strategies for each LLM: data parallelism and tensor parallelism, which is a common practice for single-node multi-GPU distributed inference~\cite{vllmParallelPractice} (pipeline parallelism is often used in multi-node multi-GPU distributed inference~\cite{vllmParallelPractice}, so we do not consider it).
We use $\mathsf{dp}$ to denote the data parallelism degree, i.e., the number of model replicates, $\mathsf{tp}$ to denote the tensor parallelism degree, i.e., the number of intra-layer model partitions.
The execution plan of a model can be represented by a tuple 
\begin{equation}
    P = (\mathsf{dp}, \mathsf{tp})
\end{equation}
$P$ is valid for a model if, according to $P$, the GPU memory is enough to store the model's weight and at least 1 sequence's KV cache (as we do not consider offloading model weights or KV cache to the CPU side in this work).
The number of GPUs required by $P$ is $\mathsf{dp}\cdot\mathsf{tp}$.

\textbf{Execution stage, its duration, and its throughput.}
The models of an application may need to be scheduled in multiple rounds, when the GPUs cannot hold all the models together or to achieve higher overall throughput (as~\Cref{fig: exec plan matters} shows).
We define an execution stage (use ``\textit{stage}'' for short) as a period of time in which a set of LLMs run concurrently with specific execution plans.
At the beginning of a stage, each model has a set of unfinished requests.
An execution stage with $k$ running models can be represented by 
\begin{equation}
    E = ((M_1, P_1), ..., (M_k, P_k))
\end{equation}

The execution stage $E$ is valid if two conditions are satisfied:
(1) for each model, its input models are either finished or in $E$ as well;
(2) each execution plan $P_i$ is valid for $M_i$ and the total number of GPUs required by the models in $E$ is smaller than the number of available GPUs.
If there is dependency among the models in $E$, then we say $E$ has \textit{model-level pipeline parallelism}.

Use $t_{M_i,P_i}$ to denote the time for $M_i$ to finish its remaining inference workload with $P_i$.
Use $t_E$ to denote the duration of $E$.
$t_E \leq \max\{t_{M_i,P_i}\}_{i=1}^k$.
If we let the models in $E$ run till the end, 
then $t_E = \max\{t_{M_i,P_i}\}_{i=1}^k$.
If we would reconsider the execution plan of each unfinished model once a model in $E$ is finished, then $t_E = \min\{t_{M_i,P_i}\}_{i=1}^k$.
Let $T_{E}$ denote the throughput of $E$, then $T_{E}=\mathsf{FLOPs}_{E}/t_{E}$, where $\mathsf{FLOPs}_{E}$ is the total FLOPs (including prefill and decode iteration FLOPs) computed for the models in $E$ during $t_{E}$.

\textbf{Application execution plan.} For a multi-LLM application, a complete execution plan is a sequence of planned execution stages, denoted by $\Phi=(E_1, ..., E_m)$.

\begin{definition}[Application Plan Search Problem]~\label{def: problem}
    Given a set of input requests and a multi-LLM application, find the best application execution plan $\Phi=(E_1, ..., E_m)$, such that (1) all the LLMs finish their inference workloads after $E_m$ ends and (2) the total stage duration $\sum_{i=1}^mt_{E_i}$ is minimized.
\end{definition}

\begin{theorem}
    The application plan search problem is NP-hard.
\end{theorem}

\textbf{Proof sketch.} This problem is NP-hard because the bin packing problem is a special case of it. 
Specifically, suppose we consider an LLM ensembling application with $n$ models (\Cref{fig:computation graphs} (a)).
Suppose each LLM $M_i$ only has one possible execution plan $P_i$, and the running time for each $M_i$ to finish its inference workload is the same, i.e., $t_{M_i, P_i}=t_{M_j, P_j}$ for $\forall i, j \in [1,n]$.
Then, each bin packing problem instance corresponds to such a problem instance.

\section{Methodology}
\begin{figure}[t]
     \centering
     \includegraphics[width=0.8\linewidth]{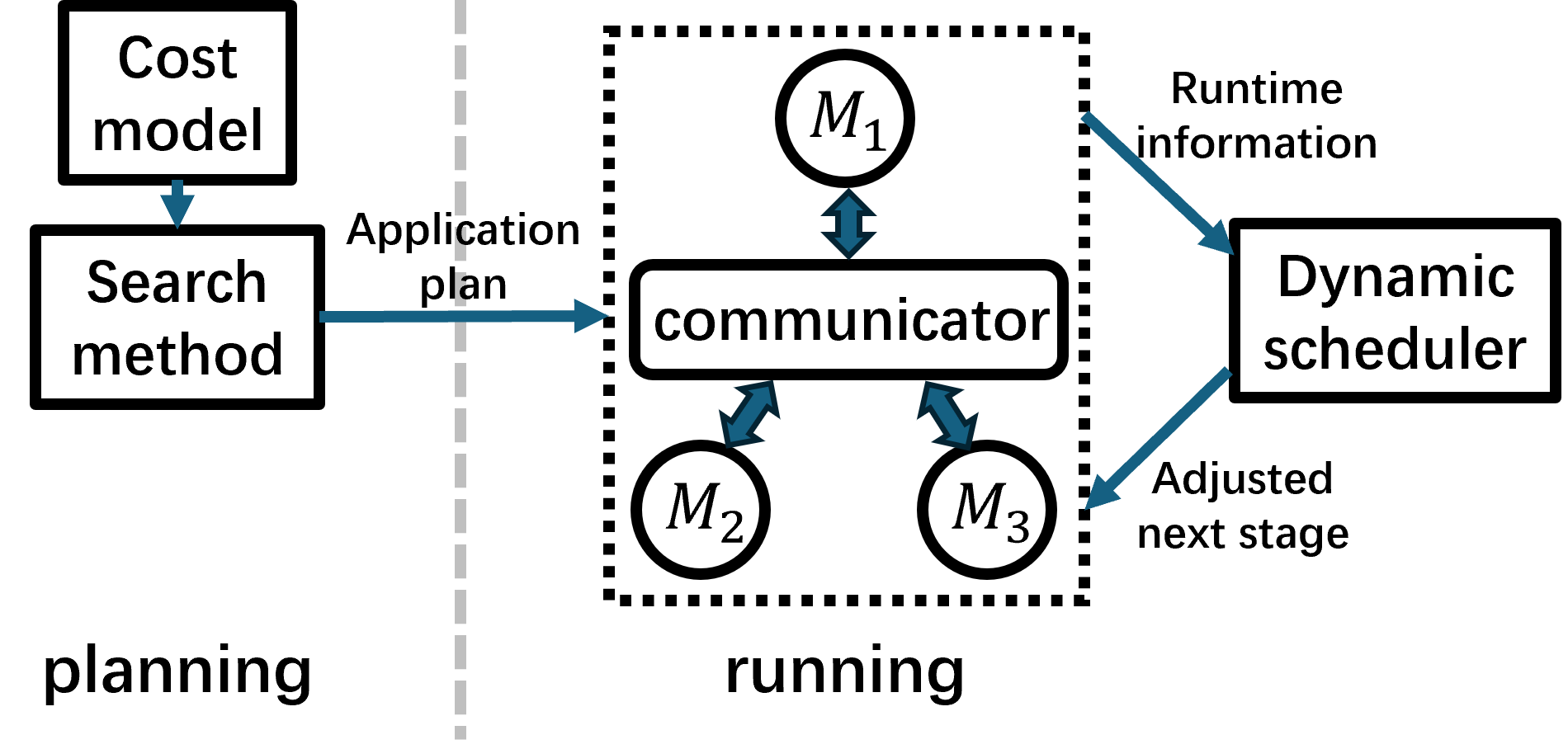}
     \caption{The framework overview.}
     \label{fig:framework}
\end{figure}

\Cref{fig:framework} shows our framework overview.
In the planning phase, to solve the problem in~\Cref{def: problem}, we construct a cost model for the LLM running time cost and perform a cost-based search for the best application execution plan.
Then in the running phase, we schedule the models according to the plan we find. 
If the actual running differs from our planning, we will dynamically adjust the execution stages using runtime information. 
Below we introduce the major components of the framework respectively.

\subsection{Cost model}
Based on the insight in~\Cref{sec: motivation and insight}, our cost model has 3 major components: the output length sampler, the request scheduling simulator, and the per-iteration latency cost model.

\textbf{Output length sampler.}
We compute the probability distribution of the output length of a specific LLM by running the LLM on a large set (e.g., 10000 requests in our experiments) of requests that do not restrict the output length implicitly in the prompt or explicitly, denoted by $F_{out}(x)$.
Suppose the maximum sequence length the LLM supports is $l_{max}$.
Then, given an input request of length $l_{in}$, the output length $l_{out}=\min(X, l_{max}-l_{in})$, $X\sim F_{out}(x)$.
If we explicitly limit the maximum output length to $y$, then $l_{out}=\min(X, y, l_{max}-l_{in})$, $X\sim F_{out}(x)$.

\textbf{Request scheduling simulator.}
To simulate the inference process of an LLM, i.e., computing the sequence lengths in each forward iteration, we need to know the sampled output lengths of the input requests and the timestamps at which each request becomes ready to be answered.
If there is no dependency among the LLMs in an execution stage, then for each of the LLMs, its input requests are all ready before the stage, and hence we can simulate the inference processes for the LLMs in any order.
Otherwise, we need to do the simulation for the LLMs in topological order.
If we fuse two models with dependency between them into one model, we will dynamically update the ready time of the input requests of the fused model during the simulation.

\textbf{Per-iteration cost model.}
Based on the insight in~\Cref{sec: motivation and insight}, we can estimate the per-iteration latency of an LLM by 
$t = t_{\mathsf{comp}} + t_{\mathsf{prep}} + t_{\mathsf{samp}}$, 
where $t_{\mathsf{comp}}, t_{\mathsf{prep}}, t_{\mathsf{samp}}$ are the time for model computation ($\mathsf{comp}$), input preparation ($\mathsf{prep}$) and output sampling ($\mathsf{samp}$), respectively.
Specifically, $t_{\mathsf{comp}}, t_{\mathsf{prep}}, t_{\mathsf{samp}}$ are all in the form of linear functions:
\begin{equation}~\label{eq: per-iter cost}
    a_{\mathsf{phase}}[B]\cdot x_{\mathsf{phase}} + b_{\mathsf{phase}}[B]
\end{equation}
where $a_{\mathsf{phase}}[B], b_{\mathsf{phase}}[B]$ denote constants specific to the $\mathsf{phase}$ ($\mathsf{comp}, \mathsf{prep}, \mathsf{samp}$) and the request number $B$, and $x_{\mathsf{phase}}$ is $\mathsf{FLOPs}$ for $\mathsf{comp}$, $B\cdot s$ for $\mathsf{prep}$, and $S$ for $\mathsf{samp}$.
We profile the per-iteration computation with different inference workloads to compute the constants in~\Cref{eq: per-iter cost}.

\textbf{Put them all together.}
For an LLM $M$ and a set of input requests $R$, we first sample the output lengths for the requests using the output length sampler, then run the request scheduling simulator to predict the running sequence information for each iteration in the whole inference process, and finally predict the latency for each iteration using the per-iteration cost model.
The summation of the latency values of all the iterations is the estimated total latency for $M$ to answer $R$.
Besides, we also need to consider the model loading time. 
As we have mentioned in~\Cref{sec: motivation and insight}, we can measure the model loading time in advance and store it in a cost table.
Note that the model loading time only needs to be considered for the newly started models in each execution stage.

\subsection{Search Method}

\begin{algorithm}[hbt!]
\caption{Greedy Search}\label{alg: greedy search}
\KwData{The computation graph $G$ of a multi-LLM application; a set of requests $R$; $N$ GPUs; a cost model $C$ }
\KwResult{An execution plan $\Phi$ for $G$}
$\Phi \gets []$\;

\While{$G$ has unfinished models}{
    $E^*\gets \{\}$; \textit{\footnotesize // the best plan group for a new stage}\textcolor{white}{\;}
    \While{True}{
        $\mathcal{M}\gets$ models in $G$ whose inputs are ready\;
        $\mathcal{P}\gets$ possible plans for $M \in\mathcal{M}$\;
        $\mathcal{E}\gets[]$\;
        \For{$P\in\mathcal{P}$}{
            \uIf{$\exists P'\in E^*$ of the same model as $P$}{
                $E\gets E^*-\{P'\}+\{P\}$\;
                \uIf{$E^*.\mathsf{\#gpu} < E.\mathsf{\#gpu}\leq N$}{
                    $\mathcal{E}.\mathsf{append}(E)$\;
                }
            }\ElseIf{$(E^*+\{P\}).\mathsf{\#gpu}\leq N$}{
                $\mathcal{E}.\mathsf{append}(E^*+\{P\})$\;
            }
        }
        $\mathcal{N}$$\gets$$[E.\mathsf{\#gpu} - E^*.\mathsf{\#gpu}|E\in\mathcal{E}]$\;
        $\mathcal{T}$$\gets$$[E.\mathsf{throughput} - E^*.\mathsf{throughput}|E\in\mathcal{E}]$\;
        \If{$\mathcal{E}$ is empty || $\mathsf{max}(\mathcal{T}) < 0$ }{
            break\;
        }
        $E^*\gets \mathcal{\mathcal{E}}[\mathsf{argmax(\mathcal{T}/\mathcal{N})}]$\; 
    }
    $\Phi.\mathsf{append}(E^*)$\;
    Update model workloads in $G$ w.r.t $E^*$\;
}
\end{algorithm}

Before we start the search, we heuristically fuse the nodes in the computation graph with self-loops into one node (the self-pointing edges are hence removed).
The search is done by a greedy method adapted from Optimus~\cite{peng2018optimus}.
\Cref{alg: greedy search} shows the pseudocode.
The main idea is that we iteratively determine the next-round execution stage until all the models can be finished (line~2).
Each execution stage $E^*$ is obtained by iteratively selecting the plan with the highest per-GPU throughput increase if we assign more GPUs to the corresponding model (line~4-23).
To avoid leaving some GPUs idle, we will end a stage once a model in the stage is finished, i.e., the running time of the stage is the shortest estimated running time of the related models.

Specifically, to obtain $E^*$, we first find the models $\mathcal{M}$ in the computation graph $G$ whose inputs are ready, i.e., its input nodes (if any) are either finished or selected in the same stage (line~5).
Then we generate possible plans for $M\in\mathcal{M}$ (line~6).
If the model $M$ of a plan $P$ has been selected by $E^*$ with plan $P'$, then if $P$ requires more GPUs than $P'$, we can consider replacing $P'$ in $E^*$ with $P$; otherwise, we consider adding $P$ to $E^*$ directly.
A new execution stage candidate $E$ is valid if the available GPUs are enough for it.
After getting all stage candidates $\mathcal{E}$, we compute the related per-GPU throughput benefits.
If there is no valid stage candidate $E$ or each candidate will only decrease the stage throughput, we stop adding more plans to the stage (line~19,20).
Then we append the current stage to the final stage list $\Phi$ and update the remaining workloads for the nodes in $G$ (line~24,25). 

\textbf{Time complexity.}
Let $V$ be the node set of $G$ and $N$ be the GPU number.
The time complexity of~\Cref{alg: greedy search} is $|V|^2N^2$.
To make the search faster, our request scheduling simulator processes different execution plans in parallel.

\subsection{Running phase}
\textbf{Communicator.} The dotted box area in~\Cref{fig:framework} illustrates how the models in a multi-LLM application are organized in our framework.
We launch one process for each node in the final computation graph to do the inference.
As there may be data dependency among nodes, we use a separate process as the communicator among the nodes to receive the node outputs, process them (e.g., applying templates), and send them to corresponding nodes upon request.

\textbf{Dynamic scheduler.} As our cost model is not perfectly accurate, the order in which the models complete all their requests may not be as estimated.
For example, for an execution stage $E_1=((M_1,P_1), (M_2, P_2), (M_3, P_3))$, 
we may plan to end $E_1$ when $M_1$ is finished (i.e., we predict $M_1$ is the first to finish among the 3 models), but in the real running process, the model that finishes first may be $M_2$.
In this case, we need to adjust the next execution stage accordingly.
Specifically, for each unfinished model $M$ in $E_1$ (i.e., $M_1$ or $M_3$), let $P$ denote its execution plan in $E_1$. 
If $E_1$ is the last stage that $M$ is involved in, then we will keep $M$ running until it is finished.
Otherwise, suppose the next stage is $E_2$. 
If $(M, P)$ is also in $E_2$ (i.e., the execution plan of $M$ is also not changed), then we will keep $M$ running, which is equivalent to scheduling $(M, P)$ from $E_2$ onto the GPUs.
If $(M, P)$ is not in $E_2$, then we will first consider scheduling the model-plan pairs in $E_2$ onto the GPUs, and then consider $(M, P)$ if there are still available GPUs, and if all GPUs are occupied, we will no longer consider the $(M, P)$ from $E_1$.
We must ensure all the model-plan pairs in a stage either have been scheduled onto the GPUs or the related models have been finished before we consider scheduling the model-plan pairs in the next stage.

When selecting GPUs to load a model, we follow the principle of minimizing model reloading costs with all the NV-link connection requirements satisfied.
For example, if there is a newly started model $M_1$ whose plan has tensor parallelism degree 2, and the NV-link in the machine only connects (GPU 0, GPU 1) and (GPU 2, GPU 3), then $M_1$ can only be loaded on GPU 0-1 or GPU 2-3. 
In this case, we may need to move some models if they occupy the GPUs required by $M_1$, and we want to choose the reloading solution with the minimum reloading cost.

\section{Experiments}
The experiments try to validate the following 4 points:
\begin{enumerate}
    \item When the inference workloads cannot saturate the GPUs, naive model scheduling solutions lead to low computation efficiency while our method can perform well.
    \item Our method is efficient to do the search.
    \item Our cost model is effective.
    \item Allowing preemption is necessary to achieve higher computation efficiency.
\end{enumerate}

We do experiments on three popular multi-LLM applications, i.e., LLM ensembling, LLM routing, and chain summary~\cite{Blender,hu2024routerbench,BooookScore} to test the performance of our method, where routing can be regarded as a special case of ensembling.
The computation graphs of these 3 applications are shown in~\Cref{fig:computation graphs}.

We use vLLM as the engine to run an LLM. 
The exaperiments are conducted on a ubuntu machine with 8 A100-80G NVIDIA GPUs (where every 2 GPUs are grouped and connected by NVLink), 2 24-core CPU, and 1510 GB RAM.

\textbf{Competitors.} We compare our method with 2 competitors:
\begin{enumerate}

    \item \textbf{Max-heuristic}: this algorithm assigns all GPUs to one LLM each time, and selects the parallelism strategy with the highest throughput for that LLM based on our cost model.

    \item \textbf{Min-heuristic}: this algorithm assigns all GPUs to as many as possible LLMs each time to maximize ``subtask'' parallelism. If there are more GPUs than the LLMs, it tries to partition the GPUs among all LLMs as evenly as possible (this method is inspired by the Min-heirustic method used in~\cite{nagrecha2023saturn}).
\end{enumerate}
These competitors all support model preemption, i.e., once a model in an execution stage finishes, all LLMs will be rescheduled according to the next execution stage plans.

\textbf{Metric.}
For each method, we compute its \textit{end-to-end running} time, which consists of 2 parts: (1) \textit{extra time}, i.e., the time to generate the execution stages, and (2) \textit{inference time}, i.e., the time to run the multi-LLM application to answer all requests.

\subsection{LLM ensembling}~\label{sec:llm ensemble}
There is no model dependency in LLM ensembling.
In this part of the experiments, we want to show how the tested methods perform as the amount of inference workload increases.

\textbf{Models.} We select the models from LLM-Blender~\cite{Blender}, including vicuna-13b-v1.5~\cite{chiang2023vicuna}, oasst-sft-4-pythia-12b-epoch-3.5~\cite{Openassistant}, alpaca-13b~\cite{alpaca}, baize-v2-13b~\cite{xu2023baize}, koala-13B-HF~\cite{Koala}, dolly-v2-12b~\cite{dolly}, mpt-7b-chat~\cite{mpt}, chatglm3-6b~\cite{glm2024chatglm}, stablelm-tuned-alpha-7b~\cite{Stability} (we do not include MOSS~\cite{moss} and Flan-T5~\cite{chung2024scaling} from LLM-Blender in the experiment because vLLM currently does not support them).

\textbf{Dataset.}
We use the MixInstruct dataset provided by LLM-Blender~\cite{Blender}.
The maximum length of the request outputs in the MixInstruct dataset~\cite{Blender} is 490 and the average is 180.
There are 2 maximum request output length limit choices in their code: 256 and 512.
Therefore, we set the limit to these 2 values to test the performance under different output lengths.
We also change the inference workload amount by setting the number of requests to values varying from 1000 to 10000.
The request input length in this dataset ranges from 5 to 127, with an average of 21.

\begin{figure}[t]%
    \centering
    \begin{subfigure}{0.45\textwidth}
        \centering
        \includegraphics[width=\linewidth]{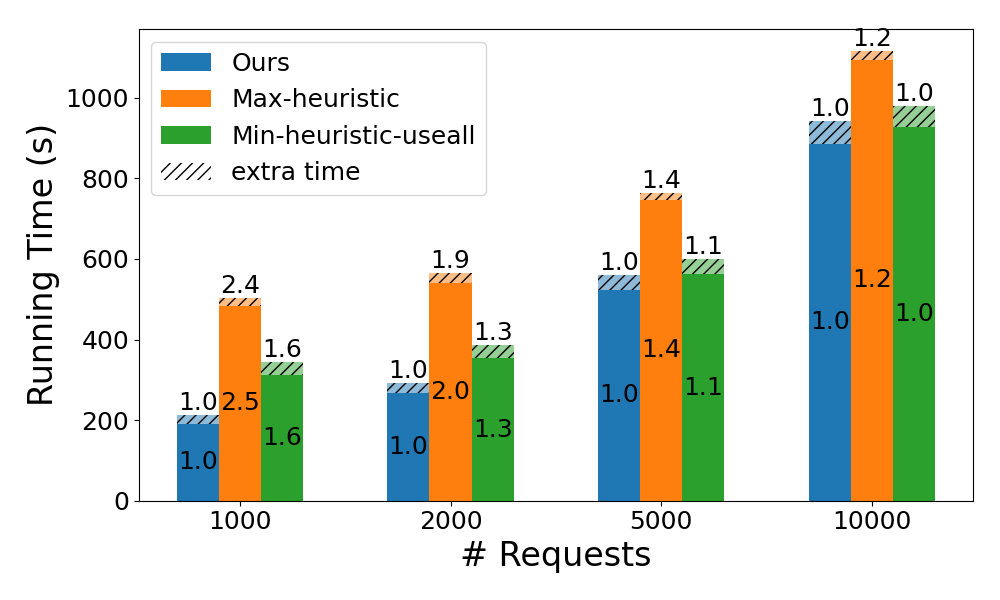} %
        \caption{Maximum output length limit = 256}
        \label{fig:blender_256}
    \end{subfigure}
    \hfill
    \begin{subfigure}{0.45\textwidth}
        \centering
        \includegraphics[width=\linewidth]{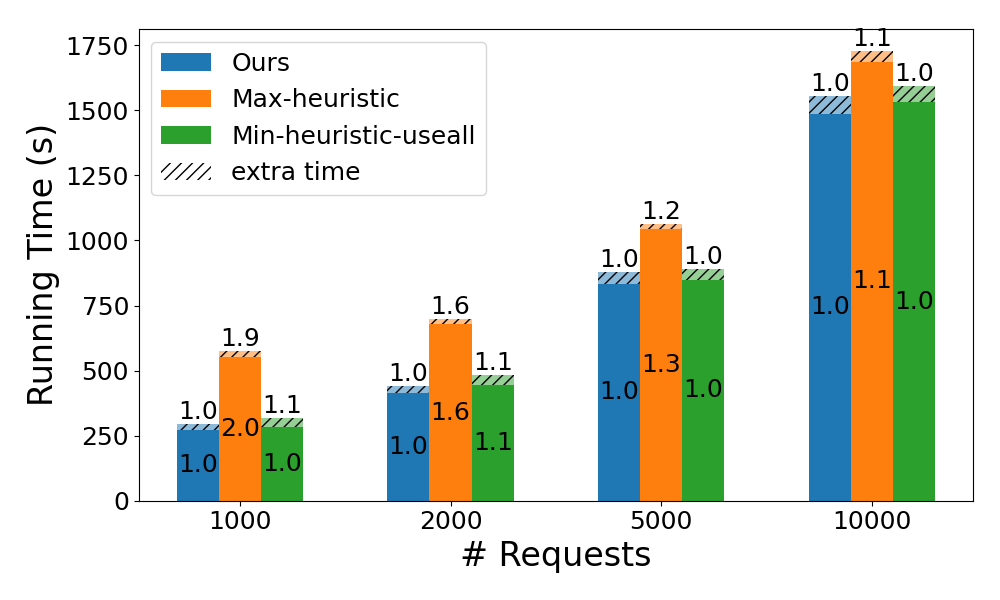} %
        \caption{Maximum output length limit = 512}
        \label{fig:blender_512}
    \end{subfigure}
    \caption{Running time VS \# Request in the LLM ensembling experiment.}
    \label{fig:blender}
\end{figure}

\textbf{Results.}
\Cref{fig:blender} shows the time to complete the LLM ensembling application of different methods under different inference workloads.
Each bar contains two parts, the inference time (the lower part) and the extra time (the upper part with slashes).
There are two values for each bar: the \textit{normalized inference time} against \ours (in the middle) and the \textit{normalized end-to-end running time} against \ours (on the top).

\textbf{\textit{Speedups}}.
In~\Cref{fig:blender}, in terms of the end-to-end running time, our method is 1.1-2.4$\times$, 1.0-1.6$\times$ better than Max-heuristic and Min-heuristic, respectively.
In terms of the LLM inference time itself, we are at most $2.5\times$ better than the competitors (when the number of requests is 1000 and the maximum output length limit is 256).

By analyzing the selected model execution plans, we know that the reasons for our speedups against the competitors are that (1) we try to assign GPUs to models to improve the overall throughput instead of each single model's throughput; (2) we take the model loading time into consideration, which ranges from 11s to 47s.
For Max-heuristic, it assigns all GPUs to each LLM every time. 
However, when the number of requests is not large enough, increasing the GPU number will not bring linear throughput improvement, and this will cause significant computation resource waste.
For instance, when the maximum output length limit is 512 and \# Requests is 1000, it takes 48s to run chatglm3-6b on 1 GPU and 32s on 8 GPUs, i.e., $8\times$ more GPUs only make the throughput $1.5\times$ higher; while when \# Requests is increased to 10000, it takes 356s to run chatglm3-6b on 1 GPU and 67s on 8 GPUs, i.e., $5.3\times$ throughput with $8\times$ GPUs.
When the model loading time is excluded, $8\times$ GPUs leads to $2.3\times$ throughput for 1000 requests while $6.6\times$ throughput for 10000 requests.
For Min-heuristic, it assigns all GPUs to the remaining LLMs in every execution stage, and this will sometimes cause unnecessary execution plan changes for LLMs.
For example, in the experiment with the maximum output length limit 256 and 1000 requests, when there are 4 LLMs unfinished, Min-heuristic assigns 2 GPUs to each LLM. After 1 LLM finishes, Min-heuristic then repartitions the GPUs to make 2 LLMs have 3 GPUs each, which requires model reloading, and 1 LLM keeps its previously assigned 2 GPUs.
However, considering the remaining request number, it would be better to keep the execution plans unchanged for the unfinished LLMs.

As the number of requests increases, the advantage of our method against the competitors becomes smaller.
One reason is that, for the models we use, when the number of requests is large enough, the number of running requests per GPU for an LLM in most of its inference time for different execution plans will be at a similar high level, so the computation throughput per GPU will be similar.
Another factor is that the ratio of the time to load models on GPUs in the overall inference process is smaller when the number of requests is larger.
Therefore, as long as we make each GPU have LLMs running on it all the time, we can achieve a good overall throughput.

\textbf{\textit{Search efficiency}}. 
The extra time of our method is $22-69$s and the proportion of it in the end-to-end running time is $4.5\%-10.5\%$ (the ratio 10.5\% corresponds to the extra time 22s when \# Requests =1000 in~\Cref{fig:blender_256}). 
In comparison, Max-heuristic and Min-heuristic require 18-41s and 31-62s extra time, respectively.
The extra time of all the 3 methods is close to each other.

\begin{table}[t]%
    \centering
    \caption{LLM selection frequency}
    \begin{tabular}{@{}lll@{}}
        \toprule
        \textbf{Model} & \textbf{\# Request} & \textbf{Ratio} \\ \midrule
        Llama-2-70b-chat-hf & 408 & 0.06 \\
        Mixtral-8x7B-Instruct-v0.1 & 1267 & 0.18 \\
        WizardLM-13B-V1.2 & 2068 & 0.30 \\
        CodeLlama-34b-Instruct-hf & 456 & 0.07 \\
        Mistral-7B-Instruct-v0.2 & 2657 & 0.39 \\ \midrule
        Total: & 6856 & 1.00 \\ \bottomrule
    \end{tabular}
    \label{tab:router req num}
\end{table}

\begin{figure}[t]
     \centering
     \includegraphics[width=0.7\linewidth]{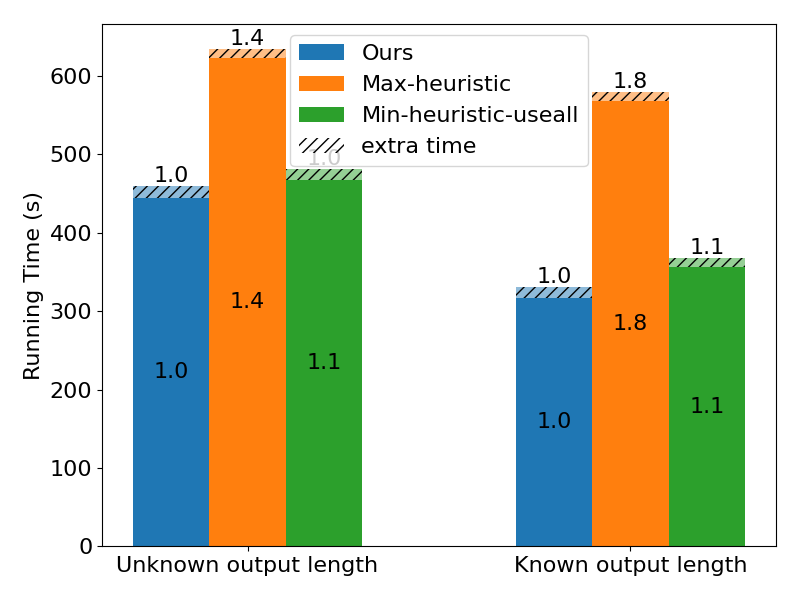}
     \caption{Running time w/o and w/ known output lengths.}
     \label{fig:router bench}
\end{figure}

\begin{figure*}[t]
    \centering
    \begin{subfigure}{0.3\textwidth}
        \centering
        \includegraphics[width=\linewidth]{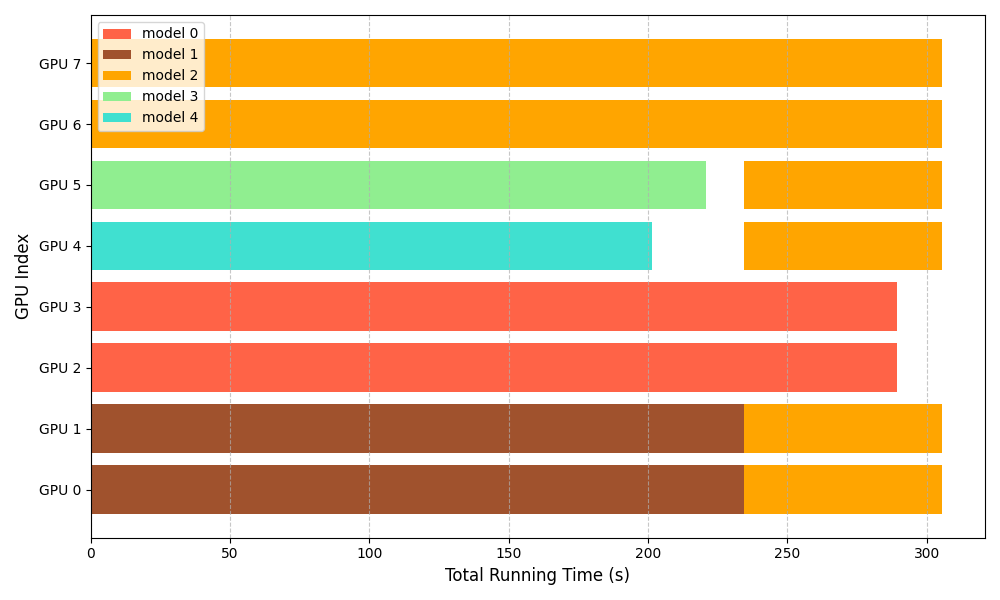} %
        \caption{Ours}
        \label{fig:schedule_router_setoutlen_greedy}
    \end{subfigure}
    \begin{subfigure}{0.3\textwidth}
        \centering
        \includegraphics[width=\linewidth]{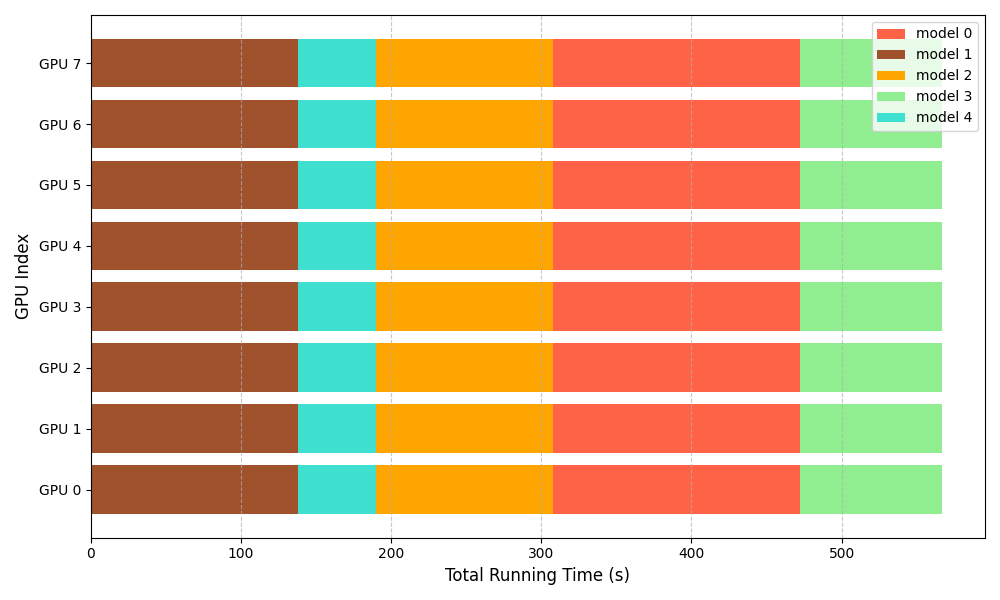} %
        \caption{Max-heuristic}
        \label{fig:schedule_router_setoutlen_max_gpu}
    \end{subfigure}
    \begin{subfigure}{0.3\textwidth}
        \centering
        \includegraphics[width=\linewidth]{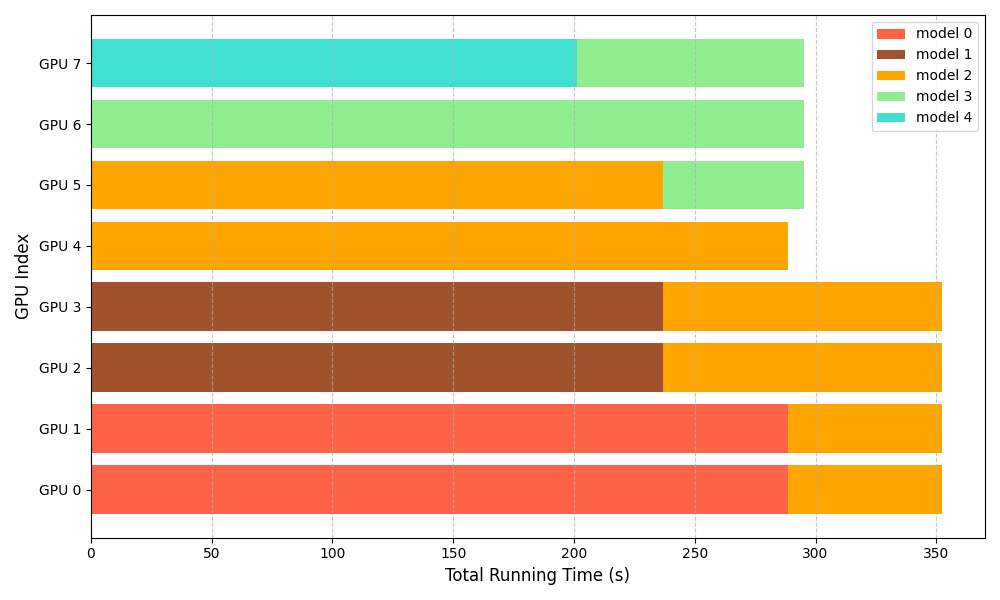} %
        \caption{Min-heuristic}
        \label{fig:schedule_router_setoutlen_min_gpu_useAll}
    \end{subfigure}
    \caption{Illustration of the LLM scheduling by different methods for LLM routing w/ known output lengths.}
    \label{fig:schedule router}
\end{figure*}

\subsection{LLM routing}
The difference between LLM routing and LLM ensembling is that in routing each request will only be routed to the most suitable LLM instead of being sent to all LLMs.
In this part of the experiments, we want to show the performance of each method when different models are assigned different numbers of requests.

\textbf{Models.} For this application, we use the models in ROUTERBENCH~\cite{hu2024routerbench}. 
Specifically, ROUTERBENCH routes requests to 11 LLMs, including both open-source LLMs and closed-source LLMs.
In our experiments, we select 5 open-source LLMs used in ROUTERBENCH, including Llama-2-70b-chat-hf~\cite{touvron2023llama}, Mixtral-8x7B-Instruct-v0.1~\cite{aggarwal2023automix}, WizardLM-13B-V1.2~\cite{xu2024wizardlm}, CodeLlama-34b-Instruct-hf~\cite{roziere2023code}, Mistral-7B-Instruct-v0.2~\cite{jiang2024identifying}.

\textbf{Dataset.} 
We use the ROUTERBENCH dataset~\cite{hu2024routerbench}.
Specifically, the ROUTERBENCH dataset stores the best-performing LLM that each request should be routed to. Therefore, for each LLM, we extract the requests to be routed to it in advance as its input. %
The number of requests for each LLM is shown in~\Cref{tab:router req num}.
The ROUTERBENCH dataset also provides the response to each request by each LLM, so we can obtain the exact output lengths from the dataset.
Therefore, we do an experiment to test different methods assuming we know the request output lengths.
The request input length of this dataset varies from 9 to 577 and the average is 310. 
The request output length of this data varies from 3 to 1585 and the average is 199.
In the situation where the request output lengths are unknown, as ROUTERBENCH does not mention the maximum output length limit when they obtain these responses, we set the output length limit to a relatively large value 4096 to test different methods.

\textbf{Results.}
\Cref{fig:router bench} shows the time to complete the LLM routing application of different methods in two situations: (1) the request output lengths are unknown and (2) we know the request output lengths (in this case, for each request, we make the LLM generate the exact number of output tokens as we set).
\Cref{fig:router bench} provides the same type of information as in~\Cref{fig:blender}.

\textbf{\textit{Speedups.}}
In terms of the end-to-end running time, our method achieves 1.4-1.8$\times$ and 1.0-1.1$\times$ speedup against Max-heuristic and Min-heuristic, respectively.
The reasons for such speedup are similar to the ones aforementioned in~\Cref{sec:llm ensemble}.
In LLM routing, even when the total number of requests is large, the LLMs with fewer requests routed to them may still have the problem of computation resource waste. 

\Cref{fig:schedule router} illustrates how different methods schedule the LLMs when the output lengths are known, which validates the reasons for our speedups.
Specifically, by comparing model 1 in~\Cref{fig:schedule_router_setoutlen_greedy} and~\Cref{fig:schedule_router_setoutlen_max_gpu}, we find that $4\times$ GPUs do not lead to $4\times$ throughput;
by comparing model 3 in~\Cref{fig:schedule_router_setoutlen_greedy} and~\Cref{fig:schedule_router_setoutlen_min_gpu_useAll}, we find that unnecessary execution plan changes waste running time.

\textbf{\textit{Search efficiency.}}
The maximum extra time of all methods is 15s, with the maximum ratio being $4.0\%$, i.e., all search methods are efficient.

\subsection{Chain summary}
As shown in~\Cref{fig:computation graphs} (d), there are two models in this application, the summary model (the $M_1$ nodes) which summarizes the document chunks, and the evaluator model (the $M_2$ node), which evaluates the quality of the generated summary in the end.
Therefore, different from LLM ensembling and routing, there is dependency between the models in the chain summary application.
Besides, due to the document length difference, the workloads of summarization models decrease progressively from left to right.

These factors can cause the GPU to idle, not merely having low computation efficiency.
In this part of experiments, we want to show the performance of each method when the model workload changes over time and appears dynamically (due to model dependency).

\textbf{Models.} 
We select vicuna-13b-v1.5 as the summary model and Llama-2-70b-chat-hf as the evaluator model.

\begin{figure}[t]
     \centering
     \includegraphics[width=0.8\linewidth]{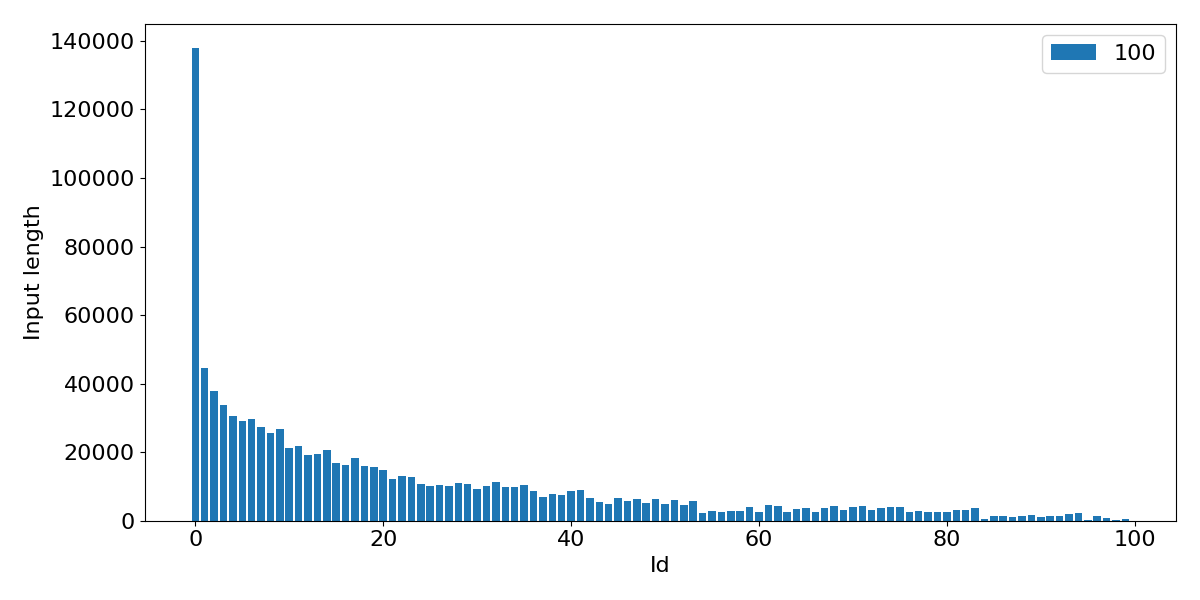}
     \caption{The lengths of the 100 sampled documents.}
     \label{fig:chain summary inp lens}
\end{figure}

\begin{figure*}[htbp]
    \centering
    \begin{subfigure}{0.3\textwidth}
        \centering
        \includegraphics[width=\linewidth]{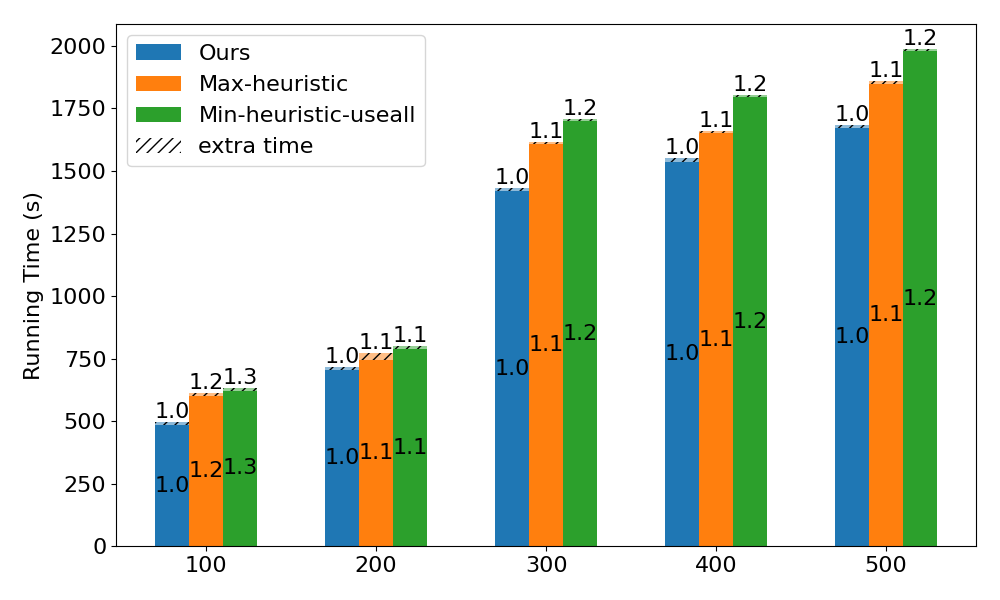} %
        \caption{Vary document number.}
        \label{fig:var reqnum}
    \end{subfigure}
    \begin{subfigure}{0.3\textwidth}
        \centering
        \includegraphics[width=\linewidth]{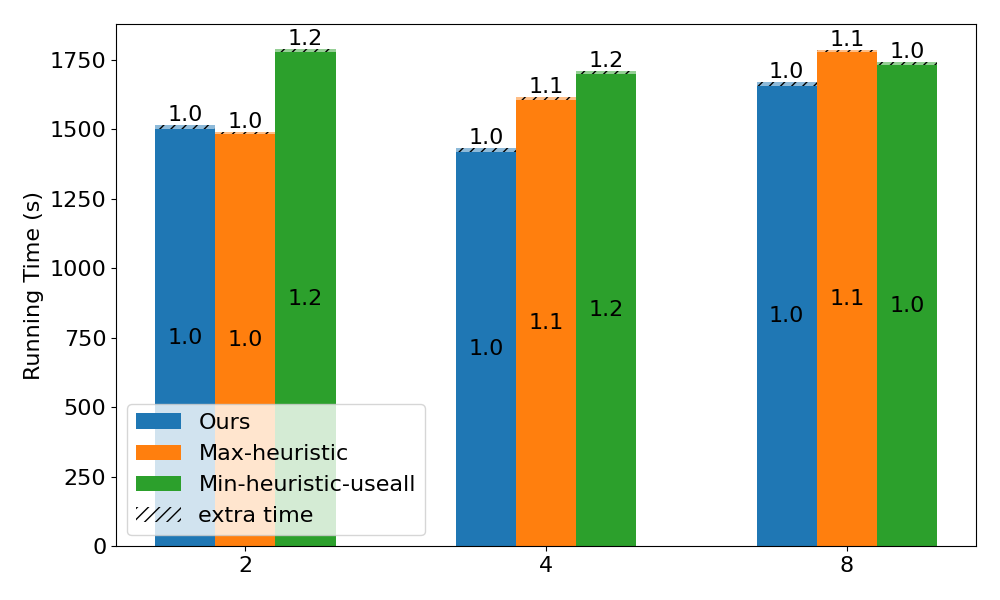} %
        \caption{Vary evaluation times.}
        \label{fig:var eval num}
    \end{subfigure}
    \begin{subfigure}{0.3\textwidth}
        \centering
        \includegraphics[width=\linewidth]{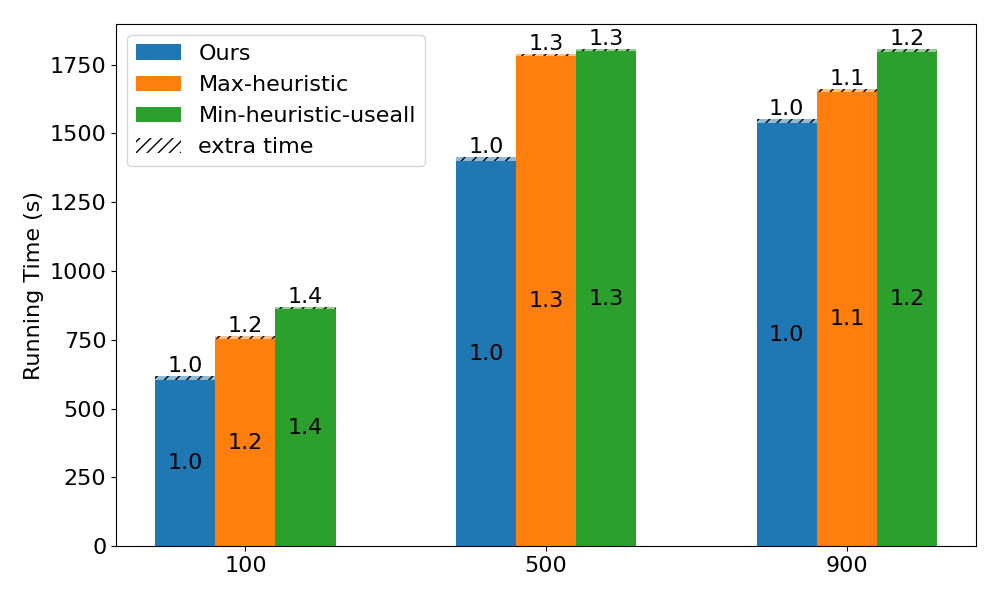} %
        \caption{Vary maximum output length.}
        \label{fig:var max outlen}
    \end{subfigure}
    \caption{Running time in different chain summary experiments.}
    \label{fig:chain summary}
\end{figure*}

\textbf{Dataset.}
We sample documents from two datasets to summarize: BOOOOKSCORE~\cite{BooookScore} and BookSum~\cite{kryscinski2022booksum}.
Each document is split into chunks of chunk size 2048 using the code in BOOOOKSCORE~\cite{BooookScore}.
We vary the sampled document number from 100 to 500.
\Cref{fig:chain summary inp lens} gives an example of 100 sampled document lengths, from which we can see that the length distribution is very skewed.
There is one extremely long document with 60 chunks, while the median length is 3 chunks.
When the sampled document number increases to 300, the largest document length is 201 chunks, while the median length is still 3 chunks.
For the evaluator model, the input is the generated summary and some instructions. The prompt template is obtained from 	
Decipherpref~\cite{hu2023decipherpref}.
We vary the number of times to evaluate a summary to simulate the scenario where a summary needs to be judged from multiple aspects.
We also test on different maximum output lengths, varying from 100 to 900.

\textbf{Results.}
\Cref{fig:chain summary} shows the time to complete the chain summary application of different methods in different dataset settings.
\Cref{fig:chain summary} provides the same type of information as in~\Cref{fig:blender,fig:router bench}.

\textbf{\textit{Speedups.}}
In terms of the end-to-end running time, our method is 1.0-1.3$\times$ and 1.0-1.4$\times$ better than Max-heuristic and Min-heuristic, respectively.
The reasons for our speedups are the same as mentioned in~\Cref{sec:llm ensemble}.
Specifically, as the summarization progresses, some GPUs become idle and can be assigned to the evaluator model, but Max-heuristic does not make use of this; while Min-heuristic would always split the GPUs evenly between the summarization model and the evaluator model, and reload the evaluator model on all GPUs after the summarization model finishes, without considering the actual inference cost.

We analyze the GPU idle time of different methods and find that Max-heuristic causes 0.9-1.6$\times$ (1.2$\times$ on average) and Min-heuristic causes 1.1-1.9$\times$ (1.5$\times$) more GPU idle time than Ours in all the tested dataset settings, which helps explain our speedups.
Note that although the GPU idle time of Max-heuristic is sometimes smaller than Ours's, the per-GPU computation efficiency when the GPUs are running is not a fixed value, so Ours is still possible to compete with Max-heuristic.
For example, when the number of evaluation times is 2 in~\Cref{fig:var eval num}, Max-heuristic has $0.9\times$ GPU idle time of Ours's. However, the summarization and the evaluation process in Max-heuristic take $1324$s and $153$s, respectively, each using 8 GPUs; while summarization in Ours takes $1479$s using 6 GPUs, and evaluation is done in parallel with summarization using 2 GPUs, with only an extra of $21$s spent after summarization finishes. From the time data we can infer that during the GPU running time, Ours has higher overall computation efficiency than Max-heuristics, and hence the total running time of the two methods is close.
Min-heuristic in this test case has $1.3\times$ more GPU idle time than Ours does, and merely summarization takes 1648s, due to only 4 GPUs assigned for it.

\textbf{\textit{Search efficiency.}}
No method's extra time exceeds $27.8$s. The maximum extra time ratio in the end-to-end running time is $3.6\%$.

\subsection{Mixed application}
In this part of the experiment, we want to compare different methods when we run more than 1 application and explore whether there are additional time-saving opportunities in this scenario.

\begin{figure}[t]
     \centering
     \includegraphics[width=1\linewidth]{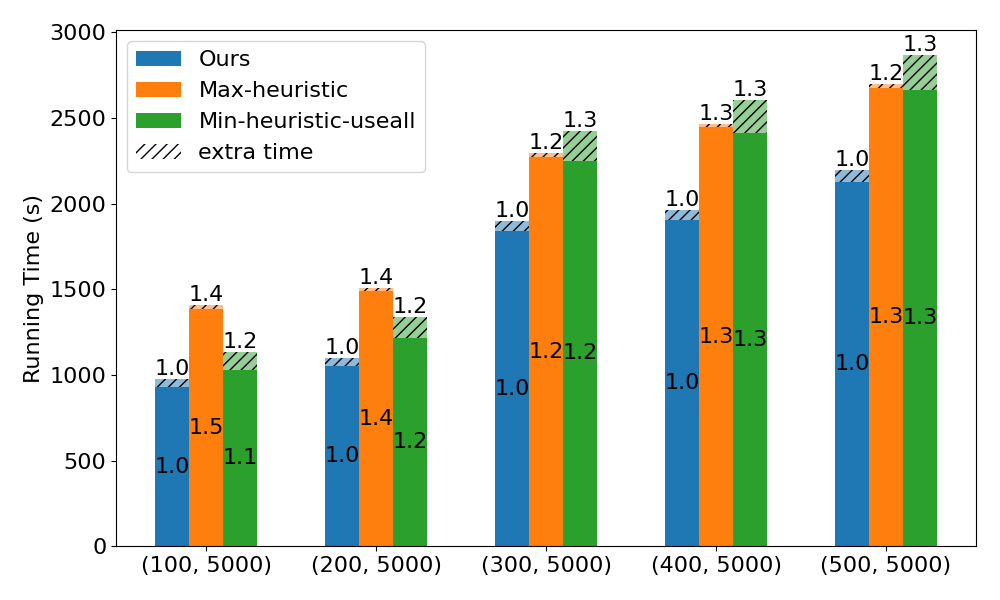}
     \caption{Running time of different request number combinations of the chain summary and the LLM ensembling.}
     \label{fig:mixed var summ reqnum}
\end{figure}

\textbf{Models.}
We run a mixture of the chain summary application and the LLM ensembling application.

\textbf{Dataset.}
We set the number of evaluation times in the chain summary to 4, the maximum output lengths of the chain summary and the LLM ensembling to 900 and 256, respectively.
We fix the number of LLM ensembling requests to 5000 and vary the number of documents to summarize from 100 to 500, to simulate different workload combinations.

\textbf{Results.}
\Cref{fig:mixed var summ reqnum} presents the same type of information as in~\Cref{fig:blender,fig:router bench,fig:chain summary}.
Each tuple on the X-axis is in the format of (\# of chain summary request, \# of LLM ensembling request).

\begin{figure*}[htbp]
    \centering
    \begin{subfigure}{0.3\textwidth}
        \centering
        \includegraphics[width=\linewidth]{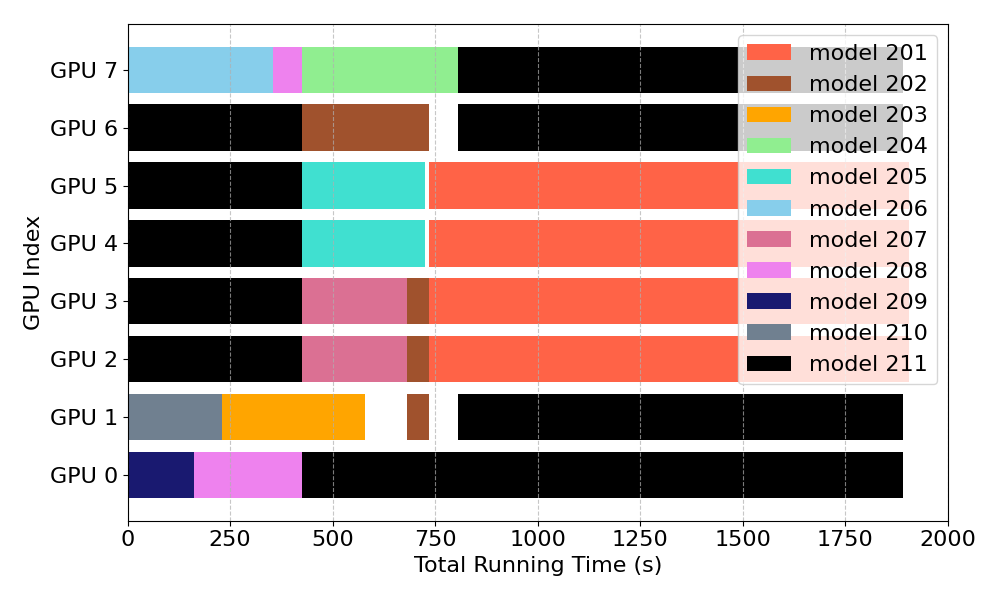} %
        \caption{Ours.}
        \label{fig:mixed ours}
    \end{subfigure}
    \begin{subfigure}{0.3\textwidth}
        \centering
        \includegraphics[width=\linewidth]{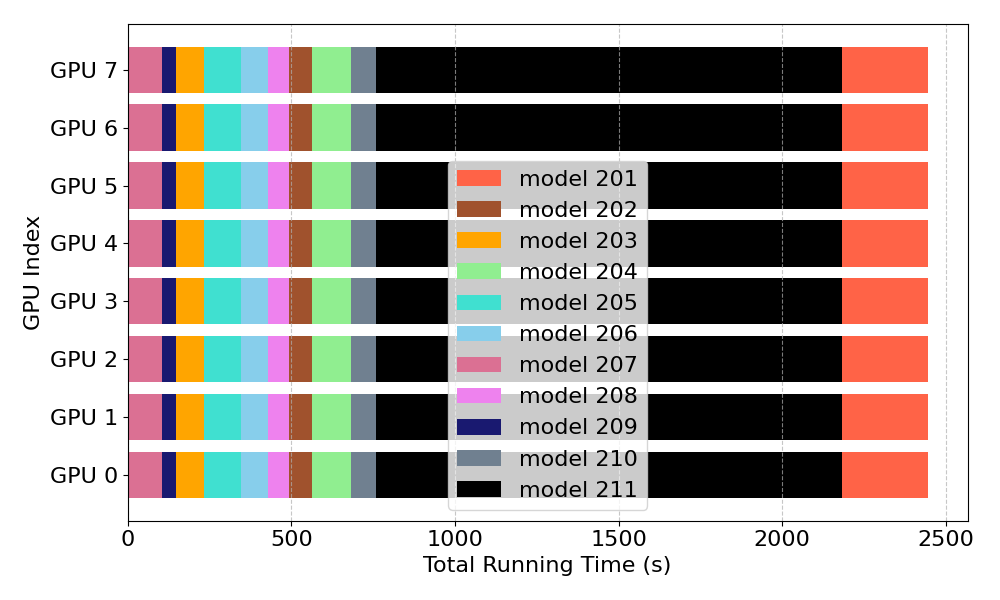} %
        \caption{Max-heuristic.}
        \label{fig:mixed max gpu}
    \end{subfigure}
    \begin{subfigure}{0.3\textwidth}
        \centering
        \includegraphics[width=\linewidth]{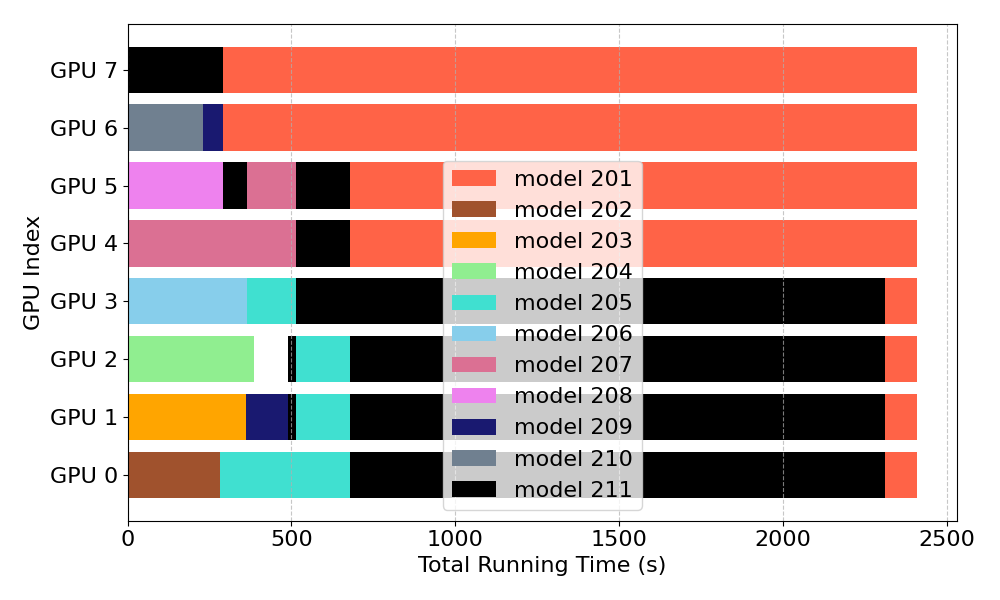} %
        \caption{Min-heuristic.}
        \label{fig:mixed min gpu}
    \end{subfigure}
    \caption{Different scheduling results in the mixed application with request number $(400,5000)$}
    \label{fig:mixed schedule}
\end{figure*}

\textbf{\textit{Speedups.}}
\Cref{fig:mixed var summ reqnum} shows that, in terms of the end-to-end running time, Ours is 1.2-1.4$\times$ better than the competitors.
By comparing~\Cref{fig:mixed var summ reqnum} against~\Cref{fig:var reqnum,fig:blender_256}, we find that using our method to run the two applications sequentially costs 1.0-1.2$\times$ longer end-to-end running time than scheduling the models by considering the two applications as a whole.
This is because the models in the LLM ensembling can make the idle GPUs busy when only a few long documents are not finished in the chain summary.
While Max-heuristic and Min-heuristic fail to do so.
Specifically, sequentially running two applications or scheduling them as a whole makes no difference for Max-heuristic.
For Min-heuristic, only when the request number tuple is $(100, 5000)$ considering the two applications as a whole brings benefit; in other cases, sequential scheduling is $0.9-1.0\times$ faster in terms of the end-to-end running time, and this is because of the increased extra time required by Min-heuristic.
In fact, in terms of the inference time, sequential scheduling is 1.0-1.2$\times$ slower than scheduling them as a whole for Min-heuristic.
\Cref{fig:mixed schedule} shows the scheduling results of the 3 methods when the request number is $(400, 5000$.
From~\Cref{fig:mixed schedule} we can see that in about the first 750s of each scheduling plan, all the models in the LLM ensembling (model 202-210) finish, but Ours additionally completes more workloads of the chain summary (model 201, 211) than the other two methods, which means we make better use of the GPU computation power in this period and helps explain our speedups against the competitors.

\textbf{\textit{Search efficiency.}}
The maximum extra time of Ours, Max-heuristic and Min-heuristic is 68s, 23s, 205s, respectively, and the maximum ratio in the end-to-end running time of them is $4.6\%, 1.7\%, 9.5\%$, respectively.
Min-heuristic needs much more extra time than the other two methods because when there are many possible execution plan combinations for one stage which evenly partition the GPUs among as many models as possible, Min-heuristic will estimate the inference time for all of them and select the one with the highest throughput.

\begin{figure}[t]
     \centering
     \includegraphics[width=1\linewidth]{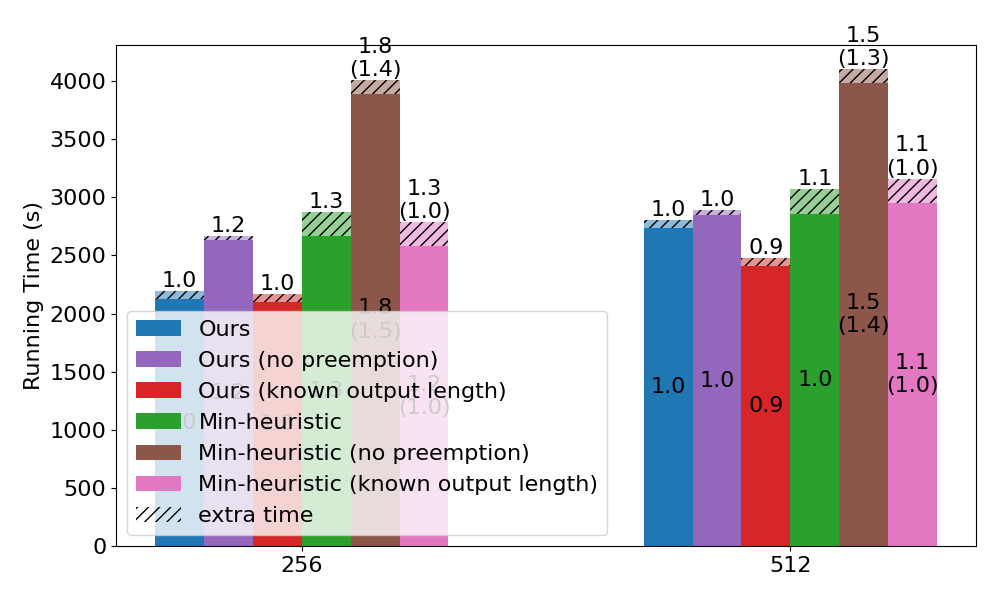}
     \caption{Running time of the method variants on the mixed application.}
     \label{fig:mixed ablation}
\end{figure}

\subsection{Ablation study}
In this part, we try to answer two questions: (1) Is introducing preemption into model scheduling beneficial? (2) How effective is our cost model for searching for model scheduling solutions compared to knowing the exact request output lengths in advance?

\textbf{Method variants.}
To answer the first question, we test the no-preemption versions of different methods, i.e., the execution plan of a model would not be changed once chosen and a running model would not be stopped once started.
To answer the second question, we replace the output lengths generated by the output sampler in the cost model with the real output lengths for each method.

\textbf{Models and Dataset.}
We run the mixed application of chain summary and LLM ensembling, with 500 requests for summarization and 5000 for ensembling.
The maximum output length limits are set to 900 and 512, respectively, and the summary evaluation is performed 4 times.

\textbf{Results.}~\Cref{fig:mixed ablation} shows the application running time of different method variants.
Note that for Max-heuristic, there is no no-preemption version, and its output-length-known version finds the same scheduling solutions as the output-length-unknown version, so we do not include Max-heuristic in~\Cref{fig:mixed ablation}.
For Min-heuristic variants, each position shows two speedup values: the top one is versus Ours, and the bracketed one is versus Min-heuristic.

\begin{figure}[t]%
    \centering
    \begin{subfigure}{0.4\textwidth}
        \centering
        \includegraphics[width=\linewidth]{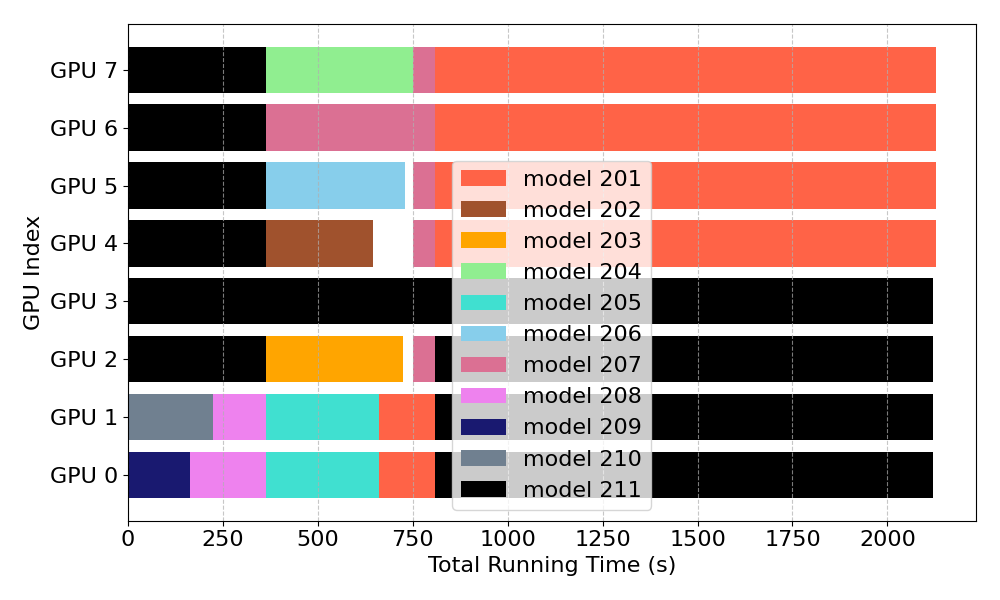} %
        \caption{Ours}
        \label{fig:mixed ablation: preempt greedy}
    \end{subfigure}
    \begin{subfigure}{0.4\textwidth}
        \centering
        \includegraphics[width=\linewidth]{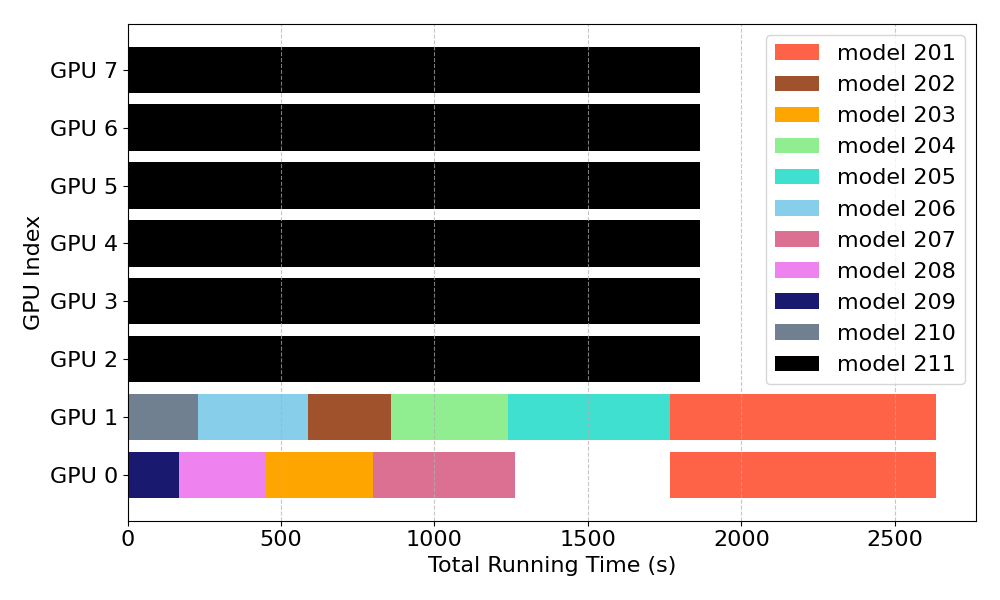} %
        \caption{Ours (no preemption)}
        \label{fig:mixed ablation no preempt greedy}
    \end{subfigure}
    \caption{Scheduling results of Ours w/ and w/o preemption when the ensembling maximum output length is 256.}
    \label{fig:mixed ablation greedy}
\end{figure}

\textbf{\textit{Preemption.}}
By comparing the no-preemption versions with the original versions of the methods in~\Cref{fig:mixed ablation}, we know that allowing preemption can bring 1.0-1.2$\times$ and 1.3-1.4$\times$ speedups in the test cases, respectively. 
As the inference process progresses, some GPUs will become idle.
Without preemption, we may not be able to utilize these GPUs for the remaining models, which significantly reduces the overall throughput.
\Cref{fig:mixed ablation greedy} illustrates the scheduling processes of Ours with and without preemption, from which we can see at the beginning model 211 (the summarization model in the chain summary) are assigned 6 GPUs. 
Without preemption, even though model 211 does not need as many GPUs later on, it continues occupying them until it finishes, forcing other models to run on only 2 GPUs. 
Although model 201 (the summary evaluator model) is started shortly before model 211 ends, it cannot use the 6 GPUs freed up afterward.
This problem is more severe for Min-heuristic. Without preemption, in both test cases, model 211 is started in the first execution stage and occupies only 1 GPU throughout its lifecycle, even if after the ensembling models finish.

\textbf{\textit{Cost model.}}
In all the application tests (including the two test cases in this part), without prior knowledge of the exact output lengths, the error ratio of the estimated inference time against the real value ranges from $6.5\%$-$38.7\%$, with an average of $25.6\%$.
With the output lengths known, for the 2 test cases of this part, the error ratio of our cost model ranges between $9.2\%$-$20.5\%$, with an average of $17.0\%$.
Even with exact output lengths, our estimation differs from the actual running time, because the cost model only includes the major computation parts (input preparation, forward computation and sampling), while other operations, like the output processing in vLLM, are not considered.
In~\Cref{fig:mixed ablation}, the output-length-known versions are 0.9-1.0$\times$ better than the output-length-unknown version, which validates that our cost model can effectively guide us to good scheduling plans even though our running time estimation is not perfectly accurate.

\section{Related Work}

\textbf{Multi-DNN inference acceleration.}
Existing works improve the multi-DNN inference efficiency by resource sharing~\cite{weng2023beware} or resource scheduling\cite{peng2018optimus}.
For example,~\cite{weng2023beware} considers GPU sharing (i.e., running multiple machine learning tasks on a single GPU) in the online task scheduling scenario and tries to address the GPU fragmentation problem.
Optimus~\cite{peng2018optimus} targets on the deep learning training jobs. 
It dynamically allocates resources and places deep learning tasks to minimize job completion time, based on online resource-performance models.
There are also works considering the online serving scenario~\cite{li2023alpaserve}.
Alpaserve~\cite{li2023alpaserve} shows that it is beneficial to use model parallelism additionally for the statistical multiplexing of multiple devices when serving multiple models, even when a model can fit into a single device, and it reduces the serving latency when there are bursty workloads.
However, these works are not designed for LLMs whose model weight size is not ignorable and may exceed the single GPU memory and whose outputs are generated auto-regressively.

\textbf{Muti-LLM inference acceleration.}
One recent popular research area about the multi-LLM inference is the multi-LLM serving system~\cite{duanmuxserve,luo2025autellix,lin2024parrot}.
MuxServe~\cite{duanmuxserve} is a flexible spatial-temporal multiplexing system
for multiple LLM serving. Its key insight is to multiplex memory resources and computation resources by colocating LLMs considering their popularity and by utilizing the LLM characteristics of prefill and decoding iterations to separate and flexibly colocate them, respectively.
Parrot~\cite{lin2024parrot} improves the end-to-end experience of the LLM-based applications by performing data flow analysis to understand the correlation across multiple LLM requests in public LLM services.
Autellix~\cite{luo2025autellix} improves the end-to-end latencies of dynamic multi-LLM agentic programs, i.e., during the execution, there are external interrupts (e.g., tool calls, human inputs).
Besides online serving, there are also works about the multi-LLM training~\cite{nagrecha2023saturn}.
Saturn~\cite{nagrecha2023saturn} reduces the model selection time by solving the joint problem of selecting parallelism, allocating resources, and scheduling.
The problem of improving the offline inference of multi-LLM application differs from the above two problems in that (1) our goal is to minimize the end-to-end running time instead of the latency of each request, and (2) inference is an auto-regressive process while training is not.

\textbf{LLM output length prediction.}
Some recent works try to predict the exact output lengths of LLMs~\cite{zheng2023response,jin2023s,hu2024inference,stojkovic2024dynamollm,fu2025efficient}.
Perception Only~\cite{zheng2023response} methods ask LLMs to generate the output length via prompting, while other methods~\cite{jin2023s,hu2024inference,stojkovic2024dynamollm} use a predictor model to predict output lengths.
\cite{fu2025efficient} trains a model to rank LLM requests by their output lengths.
Different from the model-based prediction, similar to our work, a concurrent work BlendServe~\cite{zhao2024blendserve} estimates the output lengths by sampling a subset of requests before the GPU running and assuming that the requests sharing prefixes in the prompts have a similar distribution of output length. 
While we obtain the output length distribution from a dataset different from the ones used by the application, which guides us to good overall application efficiency as well.

\section{Conclusion}
In this work, we propose a framework SamuLLM to search for the best execution plan for applications based on a sampling-then-simulation cost model and to run the application accordingly. 
The application execution plan can be adjusted dynamically during running.
Experiments show that SamuLLM can achieve 1.0-2.4$\times$ end-to-end speedups compared to the competitors.

\bibliographystyle{ACM-Reference-Format}
% \bibliography{sample}

\begin{thebibliography}{47}

%%% ====================================================================
%%% NOTE TO THE USER: you can override these defaults by providing
%%% customized versions of any of these macros before the \bibliography
%%% command.  Each of them MUST provide its own final punctuation,
%%% except for \shownote{}, \showDOI{}, and \showURL{}.  The latter two
%%% do not use final punctuation, in order to avoid confusing it with
%%% the Web address.
%%%
%%% To suppress output of a particular field, define its macro to expand
%%% to an empty string, or better, \unskip, like this:
%%%
%%% \newcommand{\showDOI}[1]{\unskip}   % LaTeX syntax
%%%
%%% \def \showDOI #1{\unskip}           % plain TeX syntax
%%%
%%% ====================================================================

\ifx \showCODEN    \undefined \def \showCODEN     #1{\unskip}     \fi
\ifx \showDOI      \undefined \def \showDOI       #1{#1}\fi
\ifx \showISBNx    \undefined \def \showISBNx     #1{\unskip}     \fi
\ifx \showISBNxiii \undefined \def \showISBNxiii  #1{\unskip}     \fi
\ifx \showISSN     \undefined \def \showISSN      #1{\unskip}     \fi
\ifx \showLCCN     \undefined \def \showLCCN      #1{\unskip}     \fi
\ifx \shownote     \undefined \def \shownote      #1{#1}          \fi
\ifx \showarticletitle \undefined \def \showarticletitle #1{#1}   \fi
\ifx \showURL      \undefined \def \showURL       {\relax}        \fi
% The following commands are used for tagged output and should be
% invisible to TeX
\providecommand\bibfield[2]{#2}
\providecommand\bibinfo[2]{#2}
\providecommand\natexlab[1]{#1}
\providecommand\showeprint[2][]{arXiv:#2}

\bibitem[\protect\citeauthoryear{Aggarwal, Madaan, Anand, Potharaju, Mishra, Zhou, Gupta, Rajagopal, Kappaganthu, Yang, et~al\mbox{.}}{Aggarwal et~al\mbox{.}}{2023}]%
        {aggarwal2023automix}
\bibfield{author}{\bibinfo{person}{Pranjal Aggarwal}, \bibinfo{person}{Aman Madaan}, \bibinfo{person}{Ankit Anand}, \bibinfo{person}{Srividya~Pranavi Potharaju}, \bibinfo{person}{Swaroop Mishra}, \bibinfo{person}{Pei Zhou}, \bibinfo{person}{Aditya Gupta}, \bibinfo{person}{Dheeraj Rajagopal}, \bibinfo{person}{Karthik Kappaganthu}, \bibinfo{person}{Yiming Yang}, {et~al\mbox{.}}} \bibinfo{year}{2023}\natexlab{}.
\newblock \showarticletitle{Automix: Automatically mixing language models}.
\newblock \bibinfo{journal}{\emph{arXiv preprint arXiv:2310.12963}} (\bibinfo{year}{2023}).
\newblock


\bibitem[\protect\citeauthoryear{Amazon}{Amazon}{2025}]%
        {Amazonprice}
\bibfield{author}{\bibinfo{person}{Amazon}.} \bibinfo{year}{2025}\natexlab{}.
\newblock \bibinfo{title}{Amazon EC2 Capacity Blocks for ML pricing}.
\newblock \bibinfo{howpublished}{\url{https://aws.amazon.com/ec2/capacityblocks/pricing/?nc1=h_ls}}.
\newblock


\bibitem[\protect\citeauthoryear{Chang, Lo, Goyal, and Iyyer}{Chang et~al\mbox{.}}{2024}]%
        {BooookScore}
\bibfield{author}{\bibinfo{person}{Yapei Chang}, \bibinfo{person}{Kyle Lo}, \bibinfo{person}{Tanya Goyal}, {and} \bibinfo{person}{Mohit Iyyer}.} \bibinfo{year}{2024}\natexlab{}.
\newblock \showarticletitle{BooookScore: {A} systematic exploration of book-length summarization in the era of LLMs}. In \bibinfo{booktitle}{\emph{The Twelfth International Conference on Learning Representations, {ICLR} 2024, Vienna, Austria, May 7-11, 2024}}. \bibinfo{publisher}{OpenReview.net}.
\newblock
\urldef\tempurl%
\url{https://openreview.net/forum?id=7Ttk3RzDeu}
\showURL{%
\tempurl}


\bibitem[\protect\citeauthoryear{Chiang, Li, Lin, Sheng, Wu, Zhang, Zheng, Zhuang, Zhuang, Gonzalez, et~al\mbox{.}}{Chiang et~al\mbox{.}}{2023}]%
        {chiang2023vicuna}
\bibfield{author}{\bibinfo{person}{Wei-Lin Chiang}, \bibinfo{person}{Zhuohan Li}, \bibinfo{person}{Ziqing Lin}, \bibinfo{person}{Ying Sheng}, \bibinfo{person}{Zhanghao Wu}, \bibinfo{person}{Hao Zhang}, \bibinfo{person}{Lianmin Zheng}, \bibinfo{person}{Siyuan Zhuang}, \bibinfo{person}{Yonghao Zhuang}, \bibinfo{person}{Joseph~E Gonzalez}, {et~al\mbox{.}}} \bibinfo{year}{2023}\natexlab{}.
\newblock \showarticletitle{Vicuna: An open-source chatbot impressing gpt-4 with 90\%* chatgpt quality}.
\newblock \bibinfo{journal}{\emph{See https://vicuna. lmsys. org (accessed 14 April 2023)}} \bibinfo{volume}{2}, \bibinfo{number}{3} (\bibinfo{year}{2023}), \bibinfo{pages}{6}.
\newblock


\bibitem[\protect\citeauthoryear{Chung, Hou, Longpre, Zoph, Tay, Fedus, Li, Wang, Dehghani, Brahma, et~al\mbox{.}}{Chung et~al\mbox{.}}{2024}]%
        {chung2024scaling}
\bibfield{author}{\bibinfo{person}{Hyung~Won Chung}, \bibinfo{person}{Le Hou}, \bibinfo{person}{Shayne Longpre}, \bibinfo{person}{Barret Zoph}, \bibinfo{person}{Yi Tay}, \bibinfo{person}{William Fedus}, \bibinfo{person}{Yunxuan Li}, \bibinfo{person}{Xuezhi Wang}, \bibinfo{person}{Mostafa Dehghani}, \bibinfo{person}{Siddhartha Brahma}, {et~al\mbox{.}}} \bibinfo{year}{2024}\natexlab{}.
\newblock \showarticletitle{Scaling instruction-finetuned language models}.
\newblock \bibinfo{journal}{\emph{Journal of Machine Learning Research}} \bibinfo{volume}{25}, \bibinfo{number}{70} (\bibinfo{year}{2024}), \bibinfo{pages}{1--53}.
\newblock


\bibitem[\protect\citeauthoryear{Conover, Hayes, Mathur, Meng, Xie, Wan, Shah, Ghodsi, Wendell, Zaharia, and Xin}{Conover et~al\mbox{.}}{2023}]%
        {dolly}
\bibfield{author}{\bibinfo{person}{Mike Conover}, \bibinfo{person}{Matt Hayes}, \bibinfo{person}{Ankit Mathur}, \bibinfo{person}{Xiangrui Meng}, \bibinfo{person}{Jianwei Xie}, \bibinfo{person}{Jun Wan}, \bibinfo{person}{Sam Shah}, \bibinfo{person}{Ali Ghodsi}, \bibinfo{person}{Patrick Wendell}, \bibinfo{person}{Matei Zaharia}, {and} \bibinfo{person}{Reynold Xin}.} \bibinfo{year}{2023}\natexlab{}.
\newblock \bibinfo{title}{Free dolly: Introducing the world’s first truly open instruction-tuned llm}.
\newblock
\newblock


\bibitem[\protect\citeauthoryear{Dam, Hong, Qiao, and Zhang}{Dam et~al\mbox{.}}{2024}]%
        {dam2024complete}
\bibfield{author}{\bibinfo{person}{Sumit~Kumar Dam}, \bibinfo{person}{Choong~Seon Hong}, \bibinfo{person}{Yu Qiao}, {and} \bibinfo{person}{Chaoning Zhang}.} \bibinfo{year}{2024}\natexlab{}.
\newblock \showarticletitle{A complete survey on llm-based ai chatbots}.
\newblock \bibinfo{journal}{\emph{arXiv preprint arXiv:2406.16937}} (\bibinfo{year}{2024}).
\newblock


\bibitem[\protect\citeauthoryear{Duan, Lu, Duanmu, Li, Zhang, Lin, Stoica, and Zhang}{Duan et~al\mbox{.}}{[n.d.]}]%
        {duanmuxserve}
\bibfield{author}{\bibinfo{person}{Jiangfei Duan}, \bibinfo{person}{Runyu Lu}, \bibinfo{person}{Haojie Duanmu}, \bibinfo{person}{Xiuhong Li}, \bibinfo{person}{Xingcheng Zhang}, \bibinfo{person}{Dahua Lin}, \bibinfo{person}{Ion Stoica}, {and} \bibinfo{person}{Hao Zhang}.} \bibinfo{year}{[n.d.]}\natexlab{}.
\newblock \showarticletitle{MuxServe: Flexible Spatial-Temporal Multiplexing for Multiple LLM Serving}. In \bibinfo{booktitle}{\emph{Forty-first International Conference on Machine Learning}}.
\newblock


\bibitem[\protect\citeauthoryear{Fu, Zhu, Su, Qiao, Stoica, and Zhang}{Fu et~al\mbox{.}}{2025}]%
        {fu2025efficient}
\bibfield{author}{\bibinfo{person}{Yichao Fu}, \bibinfo{person}{Siqi Zhu}, \bibinfo{person}{Runlong Su}, \bibinfo{person}{Aurick Qiao}, \bibinfo{person}{Ion Stoica}, {and} \bibinfo{person}{Hao Zhang}.} \bibinfo{year}{2025}\natexlab{}.
\newblock \showarticletitle{Efficient LLM Scheduling by Learning to Rank}.
\newblock \bibinfo{journal}{\emph{Advances in Neural Information Processing Systems}}  \bibinfo{volume}{37} (\bibinfo{year}{2025}), \bibinfo{pages}{59006--59029}.
\newblock


\bibitem[\protect\citeauthoryear{Geng, Gudibande, Liu, Wallace, Abbeel, Levine, and Song}{Geng et~al\mbox{.}}{2023}]%
        {Koala}
\bibfield{author}{\bibinfo{person}{Xinyang Geng}, \bibinfo{person}{Arnav Gudibande}, \bibinfo{person}{Hao Liu}, \bibinfo{person}{Eric Wallace}, \bibinfo{person}{Pieter Abbeel}, \bibinfo{person}{Sergey Levine}, {and} \bibinfo{person}{Dawn Song}.} \bibinfo{year}{2023}\natexlab{}.
\newblock \bibinfo{title}{Koala: A dialogue model for academic research}.
\newblock
\newblock


\bibitem[\protect\citeauthoryear{GLM, Zeng, Xu, Wang, Zhang, Yin, Zhang, Rojas, Feng, Zhao, et~al\mbox{.}}{GLM et~al\mbox{.}}{2024}]%
        {glm2024chatglm}
\bibfield{author}{\bibinfo{person}{Team GLM}, \bibinfo{person}{Aohan Zeng}, \bibinfo{person}{Bin Xu}, \bibinfo{person}{Bowen Wang}, \bibinfo{person}{Chenhui Zhang}, \bibinfo{person}{Da Yin}, \bibinfo{person}{Dan Zhang}, \bibinfo{person}{Diego Rojas}, \bibinfo{person}{Guanyu Feng}, \bibinfo{person}{Hanlin Zhao}, {et~al\mbox{.}}} \bibinfo{year}{2024}\natexlab{}.
\newblock \showarticletitle{Chatglm: A family of large language models from glm-130b to glm-4 all tools}.
\newblock \bibinfo{journal}{\emph{arXiv preprint arXiv:2406.12793}} (\bibinfo{year}{2024}).
\newblock


\bibitem[\protect\citeauthoryear{Hu, Huang, Xu, Chen, Xu, Chen, Feng, Wang, Wang, Bao, et~al\mbox{.}}{Hu et~al\mbox{.}}{2024b}]%
        {hu2024inference}
\bibfield{author}{\bibinfo{person}{Cunchen Hu}, \bibinfo{person}{Heyang Huang}, \bibinfo{person}{Liangliang Xu}, \bibinfo{person}{Xusheng Chen}, \bibinfo{person}{Jiang Xu}, \bibinfo{person}{Shuang Chen}, \bibinfo{person}{Hao Feng}, \bibinfo{person}{Chenxi Wang}, \bibinfo{person}{Sa Wang}, \bibinfo{person}{Yungang Bao}, {et~al\mbox{.}}} \bibinfo{year}{2024}\natexlab{b}.
\newblock \showarticletitle{Inference without interference: Disaggregate llm inference for mixed downstream workloads}.
\newblock \bibinfo{journal}{\emph{arXiv preprint arXiv:2401.11181}} (\bibinfo{year}{2024}).
\newblock


\bibitem[\protect\citeauthoryear{Hu, Bieker, Li, Jiang, Keigwin, Ranganath, Keutzer, and Upadhyay}{Hu et~al\mbox{.}}{2024a}]%
        {hu2024routerbench}
\bibfield{author}{\bibinfo{person}{Qitian~Jason Hu}, \bibinfo{person}{Jacob Bieker}, \bibinfo{person}{Xiuyu Li}, \bibinfo{person}{Nan Jiang}, \bibinfo{person}{Benjamin Keigwin}, \bibinfo{person}{Gaurav Ranganath}, \bibinfo{person}{Kurt Keutzer}, {and} \bibinfo{person}{Shriyash~Kaustubh Upadhyay}.} \bibinfo{year}{2024}\natexlab{a}.
\newblock \showarticletitle{ROUTERBENCH: A Benchmark for Multi-LLM Routing System}.
\newblock \bibinfo{journal}{\emph{arXiv preprint arXiv:2403.12031}} (\bibinfo{year}{2024}).
\newblock


\bibitem[\protect\citeauthoryear{Hu, Song, Cho, Wang, Foroosh, and Liu}{Hu et~al\mbox{.}}{2023}]%
        {hu2023decipherpref}
\bibfield{author}{\bibinfo{person}{Yebowen Hu}, \bibinfo{person}{Kaiqiang Song}, \bibinfo{person}{Sangwoo Cho}, \bibinfo{person}{Xiaoyang Wang}, \bibinfo{person}{Hassan Foroosh}, {and} \bibinfo{person}{Fei Liu}.} \bibinfo{year}{2023}\natexlab{}.
\newblock \showarticletitle{DecipherPref: Analyzing Influential Factors in Human Preference Judgments via GPT-4}. In \bibinfo{booktitle}{\emph{Proceedings of the 2023 Conference on Empirical Methods in Natural Language Processing}}. \bibinfo{pages}{8344--8357}.
\newblock


\bibitem[\protect\citeauthoryear{Jiang, Ren, and Lin}{Jiang et~al\mbox{.}}{2023}]%
        {Blender}
\bibfield{author}{\bibinfo{person}{Dongfu Jiang}, \bibinfo{person}{Xiang Ren}, {and} \bibinfo{person}{Bill~Yuchen Lin}.} \bibinfo{year}{2023}\natexlab{}.
\newblock \showarticletitle{LLM-Blender: Ensembling Large Language Models with Pairwise Ranking and Generative Fusion}. In \bibinfo{booktitle}{\emph{Proceedings of the 61st Annual Meeting of the Association for Computational Linguistics (Volume 1: Long Papers), {ACL} 2023, Toronto, Canada, July 9-14, 2023}}, \bibfield{editor}{\bibinfo{person}{Anna Rogers}, \bibinfo{person}{Jordan~L. Boyd{-}Graber}, {and} \bibinfo{person}{Naoaki Okazaki}} (Eds.). \bibinfo{publisher}{Association for Computational Linguistics}, \bibinfo{pages}{14165--14178}.
\newblock
\urldef\tempurl%
\url{https://doi.org/10.18653/V1/2023.ACL-LONG.792}
\showDOI{\tempurl}


\bibitem[\protect\citeauthoryear{Jiang}{Jiang}{2024}]%
        {jiang2024identifying}
\bibfield{author}{\bibinfo{person}{Fengqing Jiang}.} \bibinfo{year}{2024}\natexlab{}.
\newblock \emph{\bibinfo{title}{Identifying and mitigating vulnerabilities in llm-integrated applications}}.
\newblock \bibinfo{thesistype}{Master's\ thesis}. \bibinfo{school}{University of Washington}.
\newblock


\bibitem[\protect\citeauthoryear{Jin, Wu, Brooks, and Wei}{Jin et~al\mbox{.}}{2023}]%
        {jin2023s}
\bibfield{author}{\bibinfo{person}{Yunho Jin}, \bibinfo{person}{Chun{-}Feng Wu}, \bibinfo{person}{David Brooks}, {and} \bibinfo{person}{Gu{-}Yeon Wei}.} \bibinfo{year}{2023}\natexlab{}.
\newblock \showarticletitle{S\({}^{\mbox{3}}\): Increasing {GPU} Utilization during Generative Inference for Higher Throughput}. In \bibinfo{booktitle}{\emph{Advances in Neural Information Processing Systems 36: Annual Conference on Neural Information Processing Systems 2023, NeurIPS 2023, New Orleans, LA, USA, December 10 - 16, 2023}}, \bibfield{editor}{\bibinfo{person}{Alice Oh}, \bibinfo{person}{Tristan Naumann}, \bibinfo{person}{Amir Globerson}, \bibinfo{person}{Kate Saenko}, \bibinfo{person}{Moritz Hardt}, {and} \bibinfo{person}{Sergey Levine}} (Eds.).
\newblock
\urldef\tempurl%
\url{http://papers.nips.cc/paper\_files/paper/2023/hash/3a13be0c5dae69e0f08065f113fb10b8-Abstract-Conference.html}
\showURL{%
\tempurl}


\bibitem[\protect\citeauthoryear{Joel, Wu, and Fard}{Joel et~al\mbox{.}}{2024}]%
        {joel2024survey}
\bibfield{author}{\bibinfo{person}{Sathvik Joel}, \bibinfo{person}{Jie~JW Wu}, {and} \bibinfo{person}{Fatemeh~H Fard}.} \bibinfo{year}{2024}\natexlab{}.
\newblock \showarticletitle{A survey on llm-based code generation for low-resource and domain-specific programming languages}.
\newblock \bibinfo{journal}{\emph{arXiv preprint arXiv:2410.03981}} (\bibinfo{year}{2024}).
\newblock


\bibitem[\protect\citeauthoryear{Kry{\'s}ci{\'n}ski, Rajani, Agarwal, Xiong, and Radev}{Kry{\'s}ci{\'n}ski et~al\mbox{.}}{2022}]%
        {kryscinski2022booksum}
\bibfield{author}{\bibinfo{person}{Wojciech Kry{\'s}ci{\'n}ski}, \bibinfo{person}{Nazneen Rajani}, \bibinfo{person}{Divyansh Agarwal}, \bibinfo{person}{Caiming Xiong}, {and} \bibinfo{person}{Dragomir Radev}.} \bibinfo{year}{2022}\natexlab{}.
\newblock \showarticletitle{BOOKSUM: A Collection of Datasets for Long-form Narrative Summarization}. In \bibinfo{booktitle}{\emph{Findings of the Association for Computational Linguistics: EMNLP 2022}}. \bibinfo{pages}{6536--6558}.
\newblock


\bibitem[\protect\citeauthoryear{Kwon, Li, Zhuang, Sheng, Zheng, Yu, Gonzalez, Zhang, and Stoica}{Kwon et~al\mbox{.}}{2023}]%
        {kwon2023efficient}
\bibfield{author}{\bibinfo{person}{Woosuk Kwon}, \bibinfo{person}{Zhuohan Li}, \bibinfo{person}{Siyuan Zhuang}, \bibinfo{person}{Ying Sheng}, \bibinfo{person}{Lianmin Zheng}, \bibinfo{person}{Cody~Hao Yu}, \bibinfo{person}{Joseph Gonzalez}, \bibinfo{person}{Hao Zhang}, {and} \bibinfo{person}{Ion Stoica}.} \bibinfo{year}{2023}\natexlab{}.
\newblock \showarticletitle{Efficient memory management for large language model serving with pagedattention}. In \bibinfo{booktitle}{\emph{Proceedings of the 29th Symposium on Operating Systems Principles}}. \bibinfo{pages}{611--626}.
\newblock


\bibitem[\protect\citeauthoryear{LAION-AI}{LAION-AI}{2023}]%
        {Openassistant}
\bibfield{author}{\bibinfo{person}{LAION-AI}.} \bibinfo{year}{2023}\natexlab{}.
\newblock \bibinfo{title}{Open assistant}.
\newblock
\newblock
\urldef\tempurl%
\url{https:// github.com/LAION-AI/Open-Assistant}
\showURL{%
\tempurl}


\bibitem[\protect\citeauthoryear{Li, Wen, Wang, Li, Yuan, Liu, Liu, Xu, Wang, Sun, et~al\mbox{.}}{Li et~al\mbox{.}}{2024}]%
        {li2024personal}
\bibfield{author}{\bibinfo{person}{Yuanchun Li}, \bibinfo{person}{Hao Wen}, \bibinfo{person}{Weijun Wang}, \bibinfo{person}{Xiangyu Li}, \bibinfo{person}{Yizhen Yuan}, \bibinfo{person}{Guohong Liu}, \bibinfo{person}{Jiacheng Liu}, \bibinfo{person}{Wenxing Xu}, \bibinfo{person}{Xiang Wang}, \bibinfo{person}{Yi Sun}, {et~al\mbox{.}}} \bibinfo{year}{2024}\natexlab{}.
\newblock \showarticletitle{Personal llm agents: Insights and survey about the capability, efficiency and security}.
\newblock \bibinfo{journal}{\emph{arXiv preprint arXiv:2401.05459}} (\bibinfo{year}{2024}).
\newblock


\bibitem[\protect\citeauthoryear{Li, Zheng, Zhong, Liu, Sheng, Jin, Huang, Chen, Zhang, Gonzalez, et~al\mbox{.}}{Li et~al\mbox{.}}{2023}]%
        {li2023alpaserve}
\bibfield{author}{\bibinfo{person}{Zhuohan Li}, \bibinfo{person}{Lianmin Zheng}, \bibinfo{person}{Yinmin Zhong}, \bibinfo{person}{Vincent Liu}, \bibinfo{person}{Ying Sheng}, \bibinfo{person}{Xin Jin}, \bibinfo{person}{Yanping Huang}, \bibinfo{person}{Zhifeng Chen}, \bibinfo{person}{Hao Zhang}, \bibinfo{person}{Joseph~E Gonzalez}, {et~al\mbox{.}}} \bibinfo{year}{2023}\natexlab{}.
\newblock \showarticletitle{$\{$AlpaServe$\}$: Statistical multiplexing with model parallelism for deep learning serving}. In \bibinfo{booktitle}{\emph{17th USENIX Symposium on Operating Systems Design and Implementation (OSDI 23)}}. \bibinfo{pages}{663--679}.
\newblock


\bibitem[\protect\citeauthoryear{Liang, Bommasani, Lee, Tsipras, Soylu, Yasunaga, Zhang, Narayanan, Wu, Kumar, et~al\mbox{.}}{Liang et~al\mbox{.}}{2022}]%
        {liang2022holistic}
\bibfield{author}{\bibinfo{person}{Percy Liang}, \bibinfo{person}{Rishi Bommasani}, \bibinfo{person}{Tony Lee}, \bibinfo{person}{Dimitris Tsipras}, \bibinfo{person}{Dilara Soylu}, \bibinfo{person}{Michihiro Yasunaga}, \bibinfo{person}{Yian Zhang}, \bibinfo{person}{Deepak Narayanan}, \bibinfo{person}{Yuhuai Wu}, \bibinfo{person}{Ananya Kumar}, {et~al\mbox{.}}} \bibinfo{year}{2022}\natexlab{}.
\newblock \showarticletitle{Holistic evaluation of language models}.
\newblock \bibinfo{journal}{\emph{arXiv preprint arXiv:2211.09110}} (\bibinfo{year}{2022}).
\newblock


\bibitem[\protect\citeauthoryear{Lin, Han, Zhang, Yang, Yang, Chen, and Qiu}{Lin et~al\mbox{.}}{2024}]%
        {lin2024parrot}
\bibfield{author}{\bibinfo{person}{Chaofan Lin}, \bibinfo{person}{Zhenhua Han}, \bibinfo{person}{Chengruidong Zhang}, \bibinfo{person}{Yuqing Yang}, \bibinfo{person}{Fan Yang}, \bibinfo{person}{Chen Chen}, {and} \bibinfo{person}{Lili Qiu}.} \bibinfo{year}{2024}\natexlab{}.
\newblock \showarticletitle{Parrot: Efficient Serving of $\{$LLM-based$\}$ Applications with Semantic Variable}. In \bibinfo{booktitle}{\emph{18th USENIX Symposium on Operating Systems Design and Implementation (OSDI 24)}}. \bibinfo{pages}{929--945}.
\newblock


\bibitem[\protect\citeauthoryear{Luo, Shi, Cai, Zhang, Wong, Wang, Wang, Huang, Chen, Gonzalez, et~al\mbox{.}}{Luo et~al\mbox{.}}{2025}]%
        {luo2025autellix}
\bibfield{author}{\bibinfo{person}{Michael Luo}, \bibinfo{person}{Xiaoxiang Shi}, \bibinfo{person}{Colin Cai}, \bibinfo{person}{Tianjun Zhang}, \bibinfo{person}{Justin Wong}, \bibinfo{person}{Yichuan Wang}, \bibinfo{person}{Chi Wang}, \bibinfo{person}{Yanping Huang}, \bibinfo{person}{Zhifeng Chen}, \bibinfo{person}{Joseph~E Gonzalez}, {et~al\mbox{.}}} \bibinfo{year}{2025}\natexlab{}.
\newblock \showarticletitle{Autellix: An Efficient Serving Engine for LLM Agents as General Programs}.
\newblock \bibinfo{journal}{\emph{arXiv preprint arXiv:2502.13965}} (\bibinfo{year}{2025}).
\newblock


\bibitem[\protect\citeauthoryear{MosaicML}{MosaicML}{2023}]%
        {mpt}
\bibfield{author}{\bibinfo{person}{NLP~Team MosaicML}.} \bibinfo{year}{2023}\natexlab{}.
\newblock \bibinfo{title}{Introducing mpt-7b: A new standard for open-source, ly usable llms}.
\newblock
\newblock


\bibitem[\protect\citeauthoryear{Nagrecha and Kumar}{Nagrecha and Kumar}{2023}]%
        {nagrecha2023saturn}
\bibfield{author}{\bibinfo{person}{Kabir Nagrecha} {and} \bibinfo{person}{Arun Kumar}.} \bibinfo{year}{2023}\natexlab{}.
\newblock \showarticletitle{Saturn: An optimized data system for multi-large-model deep learning workloads}.
\newblock \bibinfo{journal}{\emph{Proceedings of the VLDB Endowment}} \bibinfo{volume}{17}, \bibinfo{number}{4} (\bibinfo{year}{2023}), \bibinfo{pages}{712--725}.
\newblock


\bibitem[\protect\citeauthoryear{Narayan, Chami, Orr, Arora, and R{\'e}}{Narayan et~al\mbox{.}}{2022}]%
        {narayan2022can}
\bibfield{author}{\bibinfo{person}{Avanika Narayan}, \bibinfo{person}{Ines Chami}, \bibinfo{person}{Laurel Orr}, \bibinfo{person}{Simran Arora}, {and} \bibinfo{person}{Christopher R{\'e}}.} \bibinfo{year}{2022}\natexlab{}.
\newblock \showarticletitle{Can foundation models wrangle your data?}
\newblock \bibinfo{journal}{\emph{arXiv preprint arXiv:2205.09911}} (\bibinfo{year}{2022}).
\newblock


\bibitem[\protect\citeauthoryear{Narayan, Cohen, and Lapata}{Narayan et~al\mbox{.}}{2018}]%
        {narayan2018don}
\bibfield{author}{\bibinfo{person}{Shashi Narayan}, \bibinfo{person}{Shay~B Cohen}, {and} \bibinfo{person}{Mirella Lapata}.} \bibinfo{year}{2018}\natexlab{}.
\newblock \showarticletitle{Don't give me the details, just the summary! topic-aware convolutional neural networks for extreme summarization}.
\newblock \bibinfo{journal}{\emph{arXiv preprint arXiv:1808.08745}} (\bibinfo{year}{2018}).
\newblock


\bibitem[\protect\citeauthoryear{Peng, Bao, Chen, Wu, and Guo}{Peng et~al\mbox{.}}{2018}]%
        {peng2018optimus}
\bibfield{author}{\bibinfo{person}{Yanghua Peng}, \bibinfo{person}{Yixin Bao}, \bibinfo{person}{Yangrui Chen}, \bibinfo{person}{Chuan Wu}, {and} \bibinfo{person}{Chuanxiong Guo}.} \bibinfo{year}{2018}\natexlab{}.
\newblock \showarticletitle{Optimus: an efficient dynamic resource scheduler for deep learning clusters}. In \bibinfo{booktitle}{\emph{Proceedings of the Thirteenth EuroSys Conference}}. \bibinfo{pages}{1--14}.
\newblock


\bibitem[\protect\citeauthoryear{Rajani, Tunstall, Beeching, Lambert, Rush, and Wolf}{Rajani et~al\mbox{.}}{2023}]%
        {no_robots}
\bibfield{author}{\bibinfo{person}{Nazneen Rajani}, \bibinfo{person}{Lewis Tunstall}, \bibinfo{person}{Edward Beeching}, \bibinfo{person}{Nathan Lambert}, \bibinfo{person}{Alexander~M. Rush}, {and} \bibinfo{person}{Thomas Wolf}.} \bibinfo{year}{2023}\natexlab{}.
\newblock \bibinfo{title}{No Robots}.
\newblock \bibinfo{howpublished}{\url{https://huggingface.co/datasets/HuggingFaceH4/no_robots}}.
\newblock


\bibitem[\protect\citeauthoryear{Roziere, Gehring, Gloeckle, Sootla, Gat, Tan, Adi, Liu, Sauvestre, Remez, et~al\mbox{.}}{Roziere et~al\mbox{.}}{2023}]%
        {roziere2023code}
\bibfield{author}{\bibinfo{person}{Baptiste Roziere}, \bibinfo{person}{Jonas Gehring}, \bibinfo{person}{Fabian Gloeckle}, \bibinfo{person}{Sten Sootla}, \bibinfo{person}{Itai Gat}, \bibinfo{person}{Xiaoqing~Ellen Tan}, \bibinfo{person}{Yossi Adi}, \bibinfo{person}{Jingyu Liu}, \bibinfo{person}{Romain Sauvestre}, \bibinfo{person}{Tal Remez}, {et~al\mbox{.}}} \bibinfo{year}{2023}\natexlab{}.
\newblock \showarticletitle{Code llama: Open foundation models for code}.
\newblock \bibinfo{journal}{\emph{arXiv preprint arXiv:2308.12950}} (\bibinfo{year}{2023}).
\newblock


\bibitem[\protect\citeauthoryear{Sheng, Zheng, Yuan, Li, Ryabinin, Chen, Liang, R{\'e}, Stoica, and Zhang}{Sheng et~al\mbox{.}}{2023}]%
        {sheng2023flexgen}
\bibfield{author}{\bibinfo{person}{Ying Sheng}, \bibinfo{person}{Lianmin Zheng}, \bibinfo{person}{Binhang Yuan}, \bibinfo{person}{Zhuohan Li}, \bibinfo{person}{Max Ryabinin}, \bibinfo{person}{Beidi Chen}, \bibinfo{person}{Percy Liang}, \bibinfo{person}{Christopher R{\'e}}, \bibinfo{person}{Ion Stoica}, {and} \bibinfo{person}{Ce Zhang}.} \bibinfo{year}{2023}\natexlab{}.
\newblock \showarticletitle{Flexgen: High-throughput generative inference of large language models with a single gpu}. In \bibinfo{booktitle}{\emph{International Conference on Machine Learning}}. PMLR, \bibinfo{pages}{31094--31116}.
\newblock


\bibitem[\protect\citeauthoryear{Stability-AI}{Stability-AI}{2023}]%
        {Stability}
\bibfield{author}{\bibinfo{person}{Stability-AI}.} \bibinfo{year}{2023}\natexlab{}.
\newblock \bibinfo{title}{Stablelm: Stability ai language models}.
\newblock \bibinfo{howpublished}{\url{https://github.com/ stability-AI/stableLM}}.
\newblock


\bibitem[\protect\citeauthoryear{Stojkovic, Zhang, Goiri, Torrellas, and Choukse}{Stojkovic et~al\mbox{.}}{2024}]%
        {stojkovic2024dynamollm}
\bibfield{author}{\bibinfo{person}{Jovan Stojkovic}, \bibinfo{person}{Chaojie Zhang}, \bibinfo{person}{{\'I}{\~n}igo Goiri}, \bibinfo{person}{Josep Torrellas}, {and} \bibinfo{person}{Esha Choukse}.} \bibinfo{year}{2024}\natexlab{}.
\newblock \showarticletitle{Dynamollm: Designing llm inference clusters for performance and energy efficiency}.
\newblock \bibinfo{journal}{\emph{arXiv preprint arXiv:2408.00741}} (\bibinfo{year}{2024}).
\newblock


\bibitem[\protect\citeauthoryear{Sun and Qiu}{Sun and Qiu}{2023}]%
        {moss}
\bibfield{author}{\bibinfo{person}{Tianxiang Sun} {and} \bibinfo{person}{Xipeng Qiu}.} \bibinfo{year}{2023}\natexlab{}.
\newblock \bibinfo{title}{Moss}.
\newblock \bibinfo{howpublished}{\url{https: //github.com/OpenLMLab/MOSS}}.
\newblock


\bibitem[\protect\citeauthoryear{Taori, Gulrajani, Zhang, Dubois, Li, Guestrin, Liang, and Hashimoto}{Taori et~al\mbox{.}}{2023}]%
        {alpaca}
\bibfield{author}{\bibinfo{person}{Rohan Taori}, \bibinfo{person}{Ishaan Gulrajani}, \bibinfo{person}{Tianyi Zhang}, \bibinfo{person}{Yann Dubois}, \bibinfo{person}{Xuechen Li}, \bibinfo{person}{Carlos Guestrin}, \bibinfo{person}{Percy Liang}, {and} \bibinfo{person}{Tatsunori~B. Hashimoto}.} \bibinfo{year}{2023}\natexlab{}.
\newblock \bibinfo{title}{Stanford Alpaca: An Instruction-following LLaMA model}.
\newblock \bibinfo{howpublished}{\url{https://github.com/tatsu-lab/stanford_alpaca}}.
\newblock


\bibitem[\protect\citeauthoryear{Touvron, Lavril, Izacard, Martinet, Lachaux, Lacroix, Rozi{\`e}re, Goyal, Hambro, Azhar, et~al\mbox{.}}{Touvron et~al\mbox{.}}{2023}]%
        {touvron2023llama}
\bibfield{author}{\bibinfo{person}{Hugo Touvron}, \bibinfo{person}{Thibaut Lavril}, \bibinfo{person}{Gautier Izacard}, \bibinfo{person}{Xavier Martinet}, \bibinfo{person}{Marie-Anne Lachaux}, \bibinfo{person}{Timoth{\'e}e Lacroix}, \bibinfo{person}{Baptiste Rozi{\`e}re}, \bibinfo{person}{Naman Goyal}, \bibinfo{person}{Eric Hambro}, \bibinfo{person}{Faisal Azhar}, {et~al\mbox{.}}} \bibinfo{year}{2023}\natexlab{}.
\newblock \showarticletitle{Llama: Open and efficient foundation language models}.
\newblock \bibinfo{journal}{\emph{arXiv preprint arXiv:2302.13971}} (\bibinfo{year}{2023}).
\newblock


\bibitem[\protect\citeauthoryear{vLLM}{vLLM}{2025}]%
        {vllmParallelPractice}
\bibfield{author}{\bibinfo{person}{vLLM}.} \bibinfo{year}{2025}\natexlab{}.
\newblock \bibinfo{title}{How to decide the distributed inference strategy?}
\newblock \bibinfo{howpublished}{\url{https://docs.vllm.ai/en/stable/serving/distributed_serving.html}}.
\newblock


\bibitem[\protect\citeauthoryear{Weng, Yang, Yu, Wang, Tang, Yang, and Zhang}{Weng et~al\mbox{.}}{2023}]%
        {weng2023beware}
\bibfield{author}{\bibinfo{person}{Qizhen Weng}, \bibinfo{person}{Lingyun Yang}, \bibinfo{person}{Yinghao Yu}, \bibinfo{person}{Wei Wang}, \bibinfo{person}{Xiaochuan Tang}, \bibinfo{person}{Guodong Yang}, {and} \bibinfo{person}{Liping Zhang}.} \bibinfo{year}{2023}\natexlab{}.
\newblock \showarticletitle{Beware of fragmentation: Scheduling $\{$GPU-Sharing$\}$ workloads with fragmentation gradient descent}. In \bibinfo{booktitle}{\emph{2023 USENIX Annual Technical Conference (USENIX ATC 23)}}. \bibinfo{pages}{995--1008}.
\newblock


\bibitem[\protect\citeauthoryear{Xu, Guo, Duan, and McAuley}{Xu et~al\mbox{.}}{2023}]%
        {xu2023baize}
\bibfield{author}{\bibinfo{person}{Canwen Xu}, \bibinfo{person}{Daya Guo}, \bibinfo{person}{Nan Duan}, {and} \bibinfo{person}{Julian McAuley}.} \bibinfo{year}{2023}\natexlab{}.
\newblock \showarticletitle{Baize: An open-source chat model with parameter-efficient tuning on self-chat data}.
\newblock \bibinfo{journal}{\emph{arXiv preprint arXiv:2304.01196}} (\bibinfo{year}{2023}).
\newblock


\bibitem[\protect\citeauthoryear{Xu, Sun, Zheng, Geng, Zhao, Feng, Tao, Lin, and Jiang}{Xu et~al\mbox{.}}{2024}]%
        {xu2024wizardlm}
\bibfield{author}{\bibinfo{person}{Can Xu}, \bibinfo{person}{Qingfeng Sun}, \bibinfo{person}{Kai Zheng}, \bibinfo{person}{Xiubo Geng}, \bibinfo{person}{Pu Zhao}, \bibinfo{person}{Jiazhan Feng}, \bibinfo{person}{Chongyang Tao}, \bibinfo{person}{Qingwei Lin}, {and} \bibinfo{person}{Daxin Jiang}.} \bibinfo{year}{2024}\natexlab{}.
\newblock \showarticletitle{WizardLM: Empowering large pre-trained language models to follow complex instructions}. In \bibinfo{booktitle}{\emph{The Twelfth International Conference on Learning Representations}}.
\newblock


\bibitem[\protect\citeauthoryear{Yu, Jeong, Kim, Kim, and Chun}{Yu et~al\mbox{.}}{2022}]%
        {yu2022orca}
\bibfield{author}{\bibinfo{person}{Gyeong-In Yu}, \bibinfo{person}{Joo~Seong Jeong}, \bibinfo{person}{Geon-Woo Kim}, \bibinfo{person}{Soojeong Kim}, {and} \bibinfo{person}{Byung-Gon Chun}.} \bibinfo{year}{2022}\natexlab{}.
\newblock \showarticletitle{Orca: A distributed serving system for $\{$Transformer-Based$\}$ generative models}. In \bibinfo{booktitle}{\emph{16th USENIX Symposium on Operating Systems Design and Implementation (OSDI 22)}}. \bibinfo{pages}{521--538}.
\newblock


\bibitem[\protect\citeauthoryear{Zhao, Yang, Zhu, Zheng, Kasikci, Zhou, Xing, and Stoica}{Zhao et~al\mbox{.}}{2024}]%
        {zhao2024blendserve}
\bibfield{author}{\bibinfo{person}{Yilong Zhao}, \bibinfo{person}{Shuo Yang}, \bibinfo{person}{Kan Zhu}, \bibinfo{person}{Lianmin Zheng}, \bibinfo{person}{Baris Kasikci}, \bibinfo{person}{Yang Zhou}, \bibinfo{person}{Jiarong Xing}, {and} \bibinfo{person}{Ion Stoica}.} \bibinfo{year}{2024}\natexlab{}.
\newblock \showarticletitle{BlendServe: Optimizing Offline Inference for Auto-regressive Large Models with Resource-aware Batching}.
\newblock \bibinfo{journal}{\emph{arXiv preprint arXiv:2411.16102}} (\bibinfo{year}{2024}).
\newblock


\bibitem[\protect\citeauthoryear{Zheng, Yin, Xie, Sun, Huang, Yu, Cao, Kozyrakis, Stoica, Gonzalez, et~al\mbox{.}}{Zheng et~al\mbox{.}}{2024}]%
        {zheng2024sglang}
\bibfield{author}{\bibinfo{person}{Lianmin Zheng}, \bibinfo{person}{Liangsheng Yin}, \bibinfo{person}{Zhiqiang Xie}, \bibinfo{person}{Chuyue~Livia Sun}, \bibinfo{person}{Jeff Huang}, \bibinfo{person}{Cody~Hao Yu}, \bibinfo{person}{Shiyi Cao}, \bibinfo{person}{Christos Kozyrakis}, \bibinfo{person}{Ion Stoica}, \bibinfo{person}{Joseph~E Gonzalez}, {et~al\mbox{.}}} \bibinfo{year}{2024}\natexlab{}.
\newblock \showarticletitle{Sglang: Efficient execution of structured language model programs}.
\newblock \bibinfo{journal}{\emph{Advances in Neural Information Processing Systems}}  \bibinfo{volume}{37} (\bibinfo{year}{2024}), \bibinfo{pages}{62557--62583}.
\newblock


\bibitem[\protect\citeauthoryear{Zheng, Ren, Xue, Luo, Jiang, and You}{Zheng et~al\mbox{.}}{2023}]%
        {zheng2023response}
\bibfield{author}{\bibinfo{person}{Zangwei Zheng}, \bibinfo{person}{Xiaozhe Ren}, \bibinfo{person}{Fuzhao Xue}, \bibinfo{person}{Yang Luo}, \bibinfo{person}{Xin Jiang}, {and} \bibinfo{person}{Yang You}.} \bibinfo{year}{2023}\natexlab{}.
\newblock \showarticletitle{Response length perception and sequence scheduling: An llm-empowered llm inference pipeline}.
\newblock \bibinfo{journal}{\emph{Advances in Neural Information Processing Systems}}  \bibinfo{volume}{36} (\bibinfo{year}{2023}), \bibinfo{pages}{65517--65530}.
\newblock


\end{thebibliography}
%%% -*-BibTeX-*-
%%% Do NOT edit. File created by BibTeX with style
%%% ACM-Reference-Format-Journals [18-Jan-2012].

\end{document}